\documentclass[aps,prd,onecolumn,showkeys,groupedaddress,showpacs,nofootinbib]{revtex4}
\usepackage{epsfig} \usepackage{amsmath} \usepackage{amsfonts}
\usepackage{amssymb} \usepackage{graphicx} \usepackage{colordvi}
\usepackage{psfrag}

\def\be{\begin{equation}}
\def\ee{\end{equation}}
\def\bea{\begin{eqnarray}}
\def\eea{\end{eqnarray}}

\def\a{\alpha}
\def\b{\beta}

\def\l{\lambda}

\def\Om{\Omega}
\def\pa{\partial}

\let\alp=\alpha

\renewcommand{\epsilon}{\varepsilon}

\def\beqa{\begin{eqnarray}}
\def\eeqa{\end{eqnarray}}
\def\beq{\begin{equation}}
\def\eeq{\end{equation}}

\def\pa{\partial}
\def\pr{{\it Phys. Rev.}\ }

\def\grg{{\it Gen. Relativ. Grav.}\ }

\def\mnras{{\it Mon. Not. R. Ast. Soc.}\ }

\let\alp=\alpha

\renewcommand{\epsilon}{\varepsilon}

\def\pr{{\it Phys. Rev.}\ }

\def\grg{{\it Gen. Relativ. Grav.}\ }

\def\mnras{{\it Mon. Not. R. Ast. Soc.}\ }

\def\pa{\partial}

\def\alp{\alpha}

\def\bq{{\bf q}}
\def\bqd{{\bf \dot{q}}}
\def\qd{\dot{q}}

\def\eqdef{\buildrel {\rm def} \over =}
\def\({\left(}    \def\){\right)}

\def\data{\the\day-\the\month-\the\year}

\def\frac#1#2{{#1 \over #2}}

\def\phi{\varphi}

\def\~{\approx}

\begin{document} \title{ $f(R)$ cosmology by Noether's symmetry}

\author{S. Capozziello\footnote{e\,-\,mail address:
capozziello@na.infn.it}$^{\diamond}$,\,A. De Felice\footnote{e -
mail address: antonio.defelice@uclouvain.be}$^{\natural}{}^{\ddagger}$}

\affiliation{$^{\diamond}$ Dipartimento di Scienze Fisiche,
Universit\`a di Napoli,
\\
Compl. Univ. di Monte S. Angelo,
\\Edificio
G, Via Cinthia, 80121 - Napoli, Italy
\\
 Istituto Nazionale di Fisica Nucleare, sez. di Napoli}

\affiliation{$^{\natural}$ Department of Physics and Astronomy, University of Sussex, Brighton, BN1 9QH, UK}
\affiliation{$^{\ddagger}$
Center for Particle Physics and Phenomenology (CP3)
Universit\'e catholique de Louvain
Chemin du Cyclotron, 2
B-1348 Louvain-la-Neuve,
Belgium}

\date{\today}
%----------------------------------------------------------------
\begin{abstract}
A general approach to find out exact cosmological solutions in
$f(R)$-gravity is discussed. Instead of taking into account
phenomenological models, we assume, as a physical criterium,  the
existence of Noether symmetries in the cosmological $f(R)$
Lagrangian. As a result, the presence of such symmetries selects
viable models and allow to solve  the equations of motion. We
discuss also the case in which no Noether charge is present but
general criteria can be used to achieve  solutions.
\end{abstract}


\maketitle

\section{Introduction}\label{sec1}
The recent issue to investigate alternative theories of gravity
comes out from Cosmology, Quantum Field Theory and Mach's
Principle. The initial singularity, the flatness and horizon
problems \cite{guth} point out that Standard Cosmological Model
\cite{weinberg}, based on General Relativity (GR) and Particle
Standard Model, fails in  describing the Universe at extreme
regimes. Besides, GR does not work as a fundamental theory capable
of giving a  quantum description of spacetime. Due to these
reasons and to the lack of a definitive Quantum Gravity theory,
alternative theories of gravitation have been pursued in order to
attempt, at least,  a semi-classical approach to quantization. In
particular, {\it Extended Theories of Gravity} (ETGs) face the
problem  of gravitational interaction correcting and enlarging the
Einstein theory.

The general paradigm consists in adding, into the effective
action, physically motivated higher-order curvature invariants and
non-minimally coupled scalar fields \cite{odintsov,farhoudi}.

The  interest of such an approach in early epoch cosmology is due
to the fact that ETGs can ``naturally'' reproduce inflationary behaviors
able to overcome the shortcomings of the Standard Cosmological Model
and seems also capable of matching with several observations.

From another viewpoint, the Mach Principle gives further
motivations to modify GR stating that the local inertial frame is
determined by the  average motion of distant astronomical objects
\cite{bondi}. As a consequence, the gravitational coupling can be
scale-dependent. This means that  the concept of inertia and the
Equivalence Principle have to be revised
 since there is no {\it a priori} reason to restrict the gravitational
Lagrangian to a linear function of the Ricci scalar $R$, minimally
coupled with matter \cite{brans,cimento,sciama,faraoni,Carroll:2004de,Carroll:2004hc}.

Very recently, ETGs are playing  an interesting role to describe today's
 observed Universe. In fact, the impressive amount of
good quality data of last decade seems to shed new light into the
effective picture of the Universe.  Type Ia Supernovae (SNeIa)
\cite{SNeIa}, anisotropies in the CMBR \cite{CMBR}, and matter
power spectrum derived from wide and deep galaxy surveys
\cite{LSS} represent the strongest evidences for a radical
revision of the Cosmological Standard Model also at recent epochs.

Specifically, the {\it Concordance $\Lambda$CDM Model} is showing
that baryons contribute only for $\sim 4\%$ to the total
matter\,-\,energy budget, while the  {\it cold dark matter} (CDM)
represents the bulk of the clustered large scale structures ($\sim
25\%$) and the cosmological constant $\Lambda$ plays  the role of
the so called ``dark energy'' ($\sim 70\%$) \cite{triangle}.

Although being the best fit to a wide range of data
\cite{LambdaTest}, the $\Lambda$CDM model is  affected by strong
theoretical shortcomings \cite{LambdaRev} that have motivated the
search for alternative models \cite{PR03,copeland}.

Dark energy models mainly rely on the implicit assumption that
Einstein's GR is the correct theory of gravity indeed.
Nevertheless, its validity on  large astrophysical and
cosmological scales has never been tested but only {\it assumed}
\cite{will}, and it is therefore conceivable that both cosmic
speed up and missing matter are nothing else but signals of a
breakdown of GR.  In this sense, GR could fail in giving
self-consistent pictures both at ultraviolet scales (early
universe) and at infrared scales (late universe).

Following this line of thinking, the ``minimal'' choice could be to
take into account generic functions $f(R)$ of the Ricci scalar
$R$. However, such an approach can be encompassed in the ETGs
being the minimal extension of GR.  The task for this extended
theories should be to match the data under the ``economic''
requirement that no exotic dark ingredients have to be added,
unless these are going to be found with fundamental experiments
\cite{mimicking}. This is the underlying philosophy of what are
referred to as {\it $f(R)$-gravity} (see
\cite{copeland,odirev,GRGrev}  and references therein).

Although higher order gravity theories have received much
attention in cosmology, since they are naturally able to give rise
to the accelerating expansion (both in the late and in the early
universe \cite{noi}), it is possible to demonstrate that $f(R)$
theories can also play a major role at astrophysical scales. In
fact, modifying the gravity Lagrangian  affects the gravitational
potential in the low energy limit. Provided that the modified
potential reduces to the Newtonian one on the Solar System scale,
this implication could represent an intriguing opportunity rather
than a shortcoming for $f(R)$ theories. In fact, a corrected
gravitational potential could offer the possibility to fit galaxy
rotation curves without the need of huge amounts of dark matter
\cite{noipla,mond,jcap,mnras,sobouti,salucci,mendoza}. In
addition, it is possible to work out a formal analogy between the
corrections to the Newtonian potential and the usually adopted
galaxy halo models which allow to reproduce dynamics and
observation {\it without} dark matter \cite{jcap}.

However, extending the gravitational Lagrangian could give rise to
several problems. These theories could have instabilities
\cite{DeFelice:2007zq,instabilities-f(R)} and ghost\,-\,like
behaviors \cite{ghost-f(R),DeFelice:2006pg,Calcagni:2006ye}, and
they have to be matched with the low energy limit   experiments
which fairly  test GR. Besides,  these theories should also be
compatible with early universe tests such as the formation of CMBR
anisotropies, Big Bang Nucleosynthesis \cite{DeFelice:2005bx}, and
Baryogenesis \cite{Davoudiasl:2004gf,DeFelice:2004uv}.

Actually, the debate concerning the weak field limit of
$f(R)$-gravity is far to be definitive. In the last few years,
several authors have dealt with this matter with contrasting
conclusions, in particular with respect to the Parameterized Post
Newtonian (PPN) limit \cite{ppn-tot,lgcpapers}.

In summary, it seems that the paradigm to adopt $f(R)$-gravity
leads to interesting results at cosmological, galactic and Solar
System scales  but, up to now, no definite physical criterion has
been found to {\it select} the final $f(R)$ theory (or class of
theories) capable of matching the data at all scales. Interesting
results  have been achieved in this line of thinking
\cite{mimicking,DeFelice:2007ez,Hu,Star,Odintsov1} but the
approaches are all phenomenological and are not based on some
fundamental principle as  the conservation or the invariance of
some quantity or some intrinsic symmetry of the theory.
Furthermore, as it was shown in \cite{DeFelice:2007zq},  in
alternative theories of gravity, it is important to understand the
background before exploring other bounds, such as anisotropies in
the CMBR. For this goal it is essential to try to find exact
analytical solutions for the $f(R)$ theories, and, only after
this, study more in detail the possible evolutions compatible with
our data (e.g.\ solar system and CMBR bounds).

In some sense, the situation is similar to that of dark matter: we
know very well its effect at large astrophysical scales but no
final evidence of its existence has been found, up to now, at
fundamental level. In the case of $f(R)$-gravity, we know that the
paradigm is working: in principle, the missing matter and
accelerated cosmic behavior can be addressed taking into account
gravity (in some extended version), baryons and radiation but we
do not know a specific criterion to select the final,
comprehensive theory.

In this paper, we want to address the following issues: $i)$ Is
there  some general principle capable of {\it selecting}
physically motivated $f(R)$ models? $ii)$ Can  conserved
quantities or symmetries be found in relation to specific $f(R)$
theories? $iii)$ Can such quantities, if existing,  give rise to
viable cosmological models?

In this paper, following the so called {\it Noether Symmetry
Approach} (see \cite{cimento,leandros,lambiase}, we want to seek
for viable $f(R)$ cosmological models. As we will see, the method
is twofold: from one side, the existence of symmetries allows to
solve exactly the dynamics; from the other side, the Noether {\it
charge} can always be related to some observable quantity.

The layout of the paper is the following. In Sec.\ref{sec2}, we
sketch  the dynamics of $f(R)$ gravity in the metric approach and
 derive the Friedmann-Lemaitre-Robertson-Walker (FLRW) cosmological
equations. Sec.\ref{sec3} is devoted to the general discussion of
the Noether Symmetry Approach by which it is possible to find out
conserved quantities and then symmetries which allow to exactly
solve a dynamical system. In Sec.\ref{sec4}, we apply the method
to the $f(R)$ cosmology. In Sec.\ref{sec5}, we give a detailed
summary of the exact solutions discussing them in presence or in
absence of the Noether charge. Sec.\ref{sec6} is devoted to the
discussion and the conclusions.

\section{ $f(R)$ gravity and cosmology}
\label{sec2}

The action \be S=\int d^4 x\,\sqrt{-g}\,f(R)+S_m\, , \ee describes
a theory of gravity  where $f(R)$ is a generic function of the
Ricci scalar $R$.   GR is recovered in the particular case
$f(R)=-R/16\pi G$, and $S_m$ is the action for a perfect fluid
minimally coupled with gravity \footnote{We are using the
following conventions, $\eta_{\mu\nu}= {\rm diag}(1,-1,-1,-1)$,
and $R_{\mu\nu}=R^\a{}_{\mu\a\nu}$, $c=\hbar=1$.}.

This action, in general, leads to 4th order differential equations
for the metric since the field equations are \be\label{field}
f_R\,R_{\mu\nu}-\tfrac12\,f\,g_{\mu\nu}-f_{R;\mu\nu}+g_{\mu\nu}\,\Box
f_R=-\tfrac12\,T^{m}_{\mu\nu}\, , \ee where a subscript $R$
denotes differentiation with respect to $R$ and $T^{m}_{\mu\nu}$
is the matter fluid stress-energy tensor.

Defining a {\it curvature stress\,-\,energy tensor} as
\begin{equation} \label{curva}
T^{curv}_{\mu\nu}\,=\,\frac{1}{f_{R}(R)}\left\{\frac{1}{2}g_{\mu\nu}\left[f(R)-Rf_R(R)\right]
+f_R(R)^{;\alpha\beta}(g_{\alpha\mu}g_{\beta\nu}-g_{\mu\nu}g_{\alpha\beta})
\right\}\,,
\end{equation}
Eqs.(\ref{field}) can be recast in the Einstein\,-\,like form\,:
 \begin{equation}\label{5}
G_{\mu \nu} = R_{\mu\nu}-\frac{1}{2}g_{\mu\nu}R =
T^{curv}_{\mu\nu}+T^{m}_{\mu\nu}/f_R(R)
\end{equation}
where  matter non\,-\,minimally couples to geometry through the
term $1/f_R(R)$. It is known that these theories can be mapped to
a  scalar-tensor theory. However, there are two points which
should be noticed. First, the two theories might have different
quantum descriptions, as they only coincide on the classical
solutions. Furthermore, the two theories are classically
equivalent if the Brans-Dicke parameter ($\omega_{\rm BD}$)
exactly vanishes and if the scalar field possesses a suitable
potential. This fact is related to the second point: in the
literature, the Brans-Dicke field is commonly taken as a light
scalar field for which the local gravity constraint fixes the
Brans-Dicke parameter to be greater than 40000. This bound is
usually considered when studying Brans-Dicke theories. However,
for the $f(R)$ theories, since $\omega_{\rm BD}=0$, this is not
the case, and the presence of a non-negligible potential is
essential in order to give an explicit mass to the gravitational
scalar degree of freedom. Once one has the solution $H(t)$ (and
consequently $R(t)$) for a given $f(R)$, the scalar field is
defined as $\Phi(t)=-f_R(t)$, and its potential is
$U\bigl(\Phi(t)\bigr)=R(t)\,f_R(t)-f\bigl(R(t)\bigr)$. An example
showing this link between scalar-tensor theories and $f(R)$
gravity is given in the appendix for one solution which will be
found explicitly later on.

In order to derive the cosmological equations in a FLRW metric, one
can define a canonical Lagrangian ${\cal L}={\cal L}(a,\dot{a}, R,
\dot{R})$, where  ${\cal Q}=\{a,R\}$ is the configuration space
and ${\cal TQ}=\{a,\dot{a}, R, \dot{R}\}$ is the related tangent
bundle on which ${\cal L}$ is defined. The variable $a(t)$ and
$R(t)$ are the scale factor and the  Ricci scalar in the FLRW
metric, respectively. One can use the method of the Lagrange
multipliers to set  $R$ as a constraint of the dynamics. Selecting
the suitable Lagrange multiplier and integrating by parts, the
Lagrangian ${\cal L}$ becomes canonical. In our case, we have

\be S=2\pi^2\int dt\,a^3 \left\{f(R)-\l\left[R+6\left(\frac{\ddot
a}a+\frac{\dot a^2}{a^2}+\frac \kappa{a^2}\right)\right]
-\frac{\rho_{m0}}{a^3}-\frac{\rho_{r0}}{a^4}\right\} , \ee where
$a$ is the scale factor scaled with respect to today's value (so
that $a=\tilde a/\tilde a_0$ and $a(t_0)=1$); $\rho_{m0}$ and
$\rho_{r0}$ represent  the standard amounts of dust and radiation
fluids as, for example, measured today; finally $\kappa=k/\tilde
a_0^2$, where $k=0,\pm1$. This choice for $a$, makes it
dimensionless, and it also implies that $[\kappa]=[R]=M^2$,
whereas $[f]=[\rho_{r0}]=M^4$. It is straightforward to show that,
for $f(R)=-R/16\pi G-\rho_{\Lambda0}$, one obtains the usual
Friedmann equations.

The variation with respect to $R$ of the action gives
$\l=f_R$. Therefore the previous action can be rewritten as
\be
S=2\pi^2\int dt\,a^3\left\{f-f_R\left[R+6\left(\frac{\ddot
a}a+\frac{\dot a^2}{a^2}+\frac \kappa{a^2}\right)\right]
-\frac{\rho_{m0}}{a^3}-\frac{\rho_{r0}}{a^4}\right\}\, , \ee
 and then, integrating by parts, the point-like FLRW Lagrangian is
\be {\cal L}= a^3\,(f-f_R\,R)+6\,a^2\,f_{RR}\,\dot R\,\dot a
+6\,f_R\,a\,\dot a^2-6\kappa\,f_R\,a-\rho_{m0}-\rho_{r0}/a\,
,\label{eqz0} \ee which is a canonical function of two coupled
fields, $R$ and $a$, both depending on time $t$. The total energy
$E_{\cal L}$, corresponding to the $0,0$-Einstein equation, is \be
\label{energy} E_{{\cal L}}=6\,f_{RR}\,a^2\,\dot a\,\dot R+
6\,f_R\,a\,\dot a^2- a^3\,(f-f_R\,R)
+6\kappa\,f_R\,a+\rho_{m0}+\frac{\rho_{r0}}a=0\, . \ee As we shall
see later, it is convenient to look for parametric solutions in
the form $\bigl[H(a),f\bigl(R(a)\bigr)\bigr]$, so that
$f_R=f'/R'$, where a prime denotes differentiation with respect to
the time-parameter $a$. We also have that, if $R\neq {\rm
constant}$, $f_{RR}\,\dot
R=df_R/dt=a\,H\,f_R'=a\,H\,[f''/R'-f'\,R''/{R'}^2]$, so that the
Friedmann equation can be rewritten as \be f-6 a
\left(\frac{f''}{R'}-\frac{f'\,R''}{R'^2}\right) H^2-\frac{6 f'
  \, H^2}{R'}-\left(\frac{6 \kappa
   }{a^2}+R\right)\frac{f'}{R'}=\frac{\rho_{0m}}{a^3}+\frac{\rho_{0r}}{a^4}\, .
\label{FRDA}
\ee

 The equations of motion for $a$ and $R$ are
respectively
\bea
f_{RR}\left[R+6\,H^2+6\,\frac{\ddot a}a+6\,\frac{\kappa}{a^2}\right]&=&0\label{frd1}\\
6\,f_{RRR}\,\dot R^2+ 6\,f_{RR}\,\ddot R+
6\,f_R\,H^2+12\,f_R\,\frac{\ddot a}a&=&
3\,(f-f_R\,R)-12\,f_{RR}\,H\,\dot R
-6\,f_R\,\frac{\kappa}{a^2}+\frac {\rho_{r0}}{a^4}\, ,
\label{frd2} \eea where $H\equiv \dot a/a$ is the Hubble
parameter.  Considering $R$ and $a$ as the variables,  we have,
for consistency (excluding the case $f_{RR}=0$), that $R$
coincides with the definition of the Ricci scalar in the FLRW
metric. Geometrically, this is the Euler constraint of the
dynamics. Using (\ref{frd1}),  only one of the equations
(\ref{energy}), and (\ref{frd2}) is independent because of the
Bianchi identities, as these equations correspond to the first and
second modified Einstein equations, and matter is conserved. Equivalently, after multiplying equation (\ref{frd2}) by $a^2\,\dot a$, and using (\ref{frd1}), one can integrate (\ref{frd2}) to find (\ref{energy}).
Furthermore, as we will show below, constraints on the form of the
function $f(R)$ and, consequently, solutions of the system (\ref{energy}),
(\ref{frd1}) can be achieved by asking for the existence of
Noether symmetries. Such solutions will also solve equation
(\ref{frd2}) automatically.  On the other hand, the existence of
the Noether symmetries guarantees the reduction of dynamics and
the eventual solvability of the system.

\section{The Noether Symmetry Approach}\label{sec3}

Solutions for the  dynamics given by (\ref{eqz0}) can be achieved
by selecting cyclic variables related to some Noether symmetry. In
principle, this approach allows to select $f(R)$-gravity models
compatible with the symmetry so it can be seen as a physical
criterion since the conserved quantities are a sort of {Noether
charges}. Therefore such a  criterion might be to look for those
$f(R)$ which have {\it cosmological} Noether charge. Although this
criterion somehow ``breaks'' Lorentz-invariance because we need
the FLRW background to formulate it, however Lorentz-invariance is
evidently broken in our universe by the presence of the CBMR
radiation which, by itself, fixes a preferred reference frame.

In general, the Noether Theorem states that conserved quantities
are related to the existence of cyclic variables into dynamics
\cite{arnold,marmo,morandi}.

Let ${\cal L}(q^{i}, \dot{q}^i)$ be a canonical, non-degenerate
point-like Lagrangian where \beq \label{01} \frac{\pa {\cal L
}}{\pa \lambda}=0\,;\;\;\;\;\;\;\; \mbox{det}H_{ij}\eqdef
\mbox{det} \left|\left| \frac{\pa^2 {\cal L}}{\pa
\dot{q}^i\pa\dot{q}^j}\right|\right|\neq 0\,,  \eeq with $H_{ij}$
the Hessian matrix related to ${\cal L}$ and a dot denotes
differentiation with respect to the affine parameter $\lambda$.
The dot indicates derivatives with respect to the affine parameter
$\lambda$ which, in our case, corresponds to the cosmic time $t$.
In analytical mechanics, ${\cal L}$ is of the form \beq \label{02}
{\cal L}=T({\bq},\dot{\bq})-V({\bq})\;, \eeq where $T$ and $V$ are
the ``kinetic''  and ``potential energy'' respectively. $T$ is a
positive definite quadratic form in $\dot{\bq}$.  The energy
function associated with ${\cal L}$ is \beq \label{03} E_{\cal L
}\equiv\frac{\pa {\cal L}}{\pa \qd^{i}}\qd^i-{\cal L}\,, \eeq
which is the total energy $T+V$. In any case, $E_{\cal L }$ is a
constant of  motion. Since our cosmological problem has a finite
number of degrees of freedom,  we are going to consider only
point-transformations. Any invertible  transformation of the
``generalized positions'' $Q^{i}=Q^{i}({\bq})$ induces a
transformation of the ``generalized velocities'' such that \beq
\label{04} \dot{Q}^i({\bq})=\frac{\pa Q^i}{\pa q^j}\qd^j\;; \eeq
the matrix ${\cal J}=|| \pa Q^i/\pa q^j ||$ is the Jacobian of the
transformation on the positions, and it is assumed to be nonzero.
The Jacobian $\widetilde{{\cal J}}$ of the induced transformation
is easily derived and ${\cal J}\neq 0\rightarrow \widetilde{{\cal
J}}\neq 0$. In general, this condition is not satisfied in the
whole space but only in the neighbor of a point. It is a  local
transformation.

A point transformation $Q^{i}=Q^{i}(\bq)$ can depend on one (or
more than one) parameter. As starting point, we can assume that a
point transformation depends on a parameter $\epsilon$, i.e.
$Q^{i}=Q^{i}(\bq,\epsilon)$, and that it gives rise to a
one-parameter Lie group. For infinitesimal values of $\epsilon$,
the transformation is then generated by a vector field: for
instance,  $\pa/\pa x$ is a translation along the $x$ axis,
$x(\pa/\pa y)-y(\pa/\pa x)$ is a rotation around the $z$ axis and
so on. The induced transformation (\ref{04}) is then represented
by \beq \label{05} {\bf X}=\alp^{i}({\bq})\frac{\pa}{\pa q^{i}}+
\left(\frac{d}{d\lambda}\alp^{i}({\bq})\right)\frac{\pa}{\pa
\qd^i}\;. \eeq ${\bf X}$ is called the ``complete lift'' of ${\bf
X}$ \cite{morandi}. A function $F(\bq, \bqd)$ is invariant under
the transformation  ${\bf X}$ if \beq \label{06} L_{{\bf
X}}F\eqdef\alp^{i}({\bq})\frac{\pa F}{\pa q^{i}}+
\left(\frac{d}{d\lambda}\alp^{i}({\bq})\right)\frac{\pa F}{\pa
\qd^i}\,=\,0\;, \eeq where $L_{{{\bf X}}}F$ is the Lie derivative
of $F$. Specifically, if $L_{{{\bf X}}}{\cal L}=0$, ${\bf X}$ is a
{\it symmetry} for the dynamics derived by ${\cal L}$. 
As we shall see later on, we will look for a sufficient condition on the form of $f(R)$ in our Lagrangian, which allows $L_{{{\bf X}}}{\cal L}=0$ to vanish.

Let us consider now a Lagrangian ${\cal L}$ and its Euler-Lagrange
equations \beq \label{07} \frac{d}{d\lambda}\frac{\pa {\cal L
}}{\pa\qd^{j}}-\frac{\pa {\cal L}}{\pa q^{j}}=0\,. \eeq Let us
consider also the vector field (\ref{05}). Contracting (\ref{07})
with the $\alpha^{i}$'s gives \beq \label{06a} \alp^{j}\left(
\frac{d}{d\lambda}\frac{\pa {\cal L}}{\pa \qd^j}- \frac{\pa {\cal
L }}{\pa q^j}\right)=0\,. \eeq Being \beq \label{06b}
\alp^{j}\frac{d}{d\lambda}\frac{\pa {\cal L}}{\pa \qd^j}=
\frac{d}{d\lambda}\left(\alp^j\frac{\pa {\cal L}}{\pa \qd
^j}\right)- \left(\frac{d \alp^j}{d\lambda}\right)\frac{\pa {\cal
L }}{\pa \qd ^j}\,, \eeq from (\ref{06a}), we obtain \beq
\label{08} \frac{d}{d\lambda}\left(\alp^{i}\frac{\pa {\cal L}}{\pa
\qd^i} \right)=L_{\bf X}{\cal L}\,. \eeq The immediate consequence
is the {\it Noether Theorem} which states:

If $L_{\bf X}{\cal L}=0$, then the function \beq \label{09}
\Sigma_{0}=\alp^{i}\frac{\pa {\cal L }}{\pa \qd^i} \,, \eeq is a
constant of motion.

Some comments are necessary at this point.  Eq.(\ref{09}) can be
expressed independently of coordinates as a contraction of ${\bf
X}$ by a Cartan one-form \beq \label{09a} \theta_{\cal L} \eqdef
\frac{\pa {\cal L}}{\pa \qd^i}dq^i \; . \eeq For a generic vector
field $ {\bf Y} = y^i \pa / \pa x^i $, and one-form $\beta =
\beta_i d x^i $, we have, by definition, $ {\displaystyle i_{\bf
Y} \beta = y^i \beta_i} $. Thus Eq.(\ref{09}) can be written as
\beq \label{09b} i_{\bf X} \theta _{\cal L} = \Sigma_{0} \; . \eeq
By a point-transformation, the vector field ${\bf X}$ becomes \beq
\label{09c} \widetilde{{\bf X}} = (i_{\bf X} d Q^k) \frac{\pa}{\pa
Q^k} +
     \left( \frac{d}{d\lambda} (i_x d Q^k)\right) \frac{\pa}{\pa \dot{Q}^k} \; .
\eeq We see that $\widetilde{{\bf X}}'$ is still the lift of a
vector field defined on the ``space of positions.'' If ${\bf X}$ is
a symmetry and we choose a point transformation  such  that \beq
\label{010} i_{\bf X} dQ^1 = 1 \; ; \;\;\; i_{\bf X} dQ^i = 0
\;\;\; i \neq 1 \; , \eeq we get \beq \label{010a} \widetilde{{\bf
X}} = \frac{\pa}{\pa Q^1} \;;\;\;\;\;  \frac{\pa {\cal L}}{\pa
Q^1} = 0 \; . \eeq Thus $Q^1$ is a cyclic coordinate and the
dynamics results {\it reduced} \cite{arnold,marmo}.

Furthermore, the change of coordinates given by (\ref{010}) is not
unique and then  a clever choice could be very important.  In
general, the solution of Eq.(\ref{010}) is not  defined on the
whole space. It is  local in the sense explained above. Besides,
it is possible that more than one ${\bf X}$ is found, e.g.  ${\bf
X}_1$, ${\bf X}_2$. If they commute, i.e. $ [{\bf X}_1, {\bf X}_2]
= 0 $, then it is possible to obtain two cyclic coordinates by
solving the system \beq i_{\bf {X_{1}}} dQ^1 = 1; \,  i_{\bf
{X_{2}}} dQ^2 = 1; \, i_{\bf {X_{1}}} dQ^i = 0;\,i \neq 1;\,
i_{\bf {X_{2}}} dQ^i = 0; \, i \neq 2\,. \eeq The transformed
fields will be $\pa/\pa Q^{1}$, $\pa/\pa Q^{2}$. If they do not
commute, this procedure is clearly not applicable, since
commutation relations are preserved by diffeomorphisms. If the
relation  ${\bf X}_3 = [{\bf X}_1, {\bf X}_2]$ holds,  also  ${\bf
X}_3$ is a symmetry, being $L_{\bf{X_{3}}}{\cal
L}=L_{\bf{X_{1}}}L_{\bf{X_{2}}}{\cal L}-
L_{\bf{X_{2}}}L_{\bf{X_{1}}}{\cal L}=0$. If ${\bf X_{3}}$ is
independent of ${\bf X_{1}}$, ${\bf X_{2}}$, we can go on until
the vector fields close the Lie algebra. The usual approach to
this situation is to make a Legendre transformation, going to the
Hamiltonian formalism, and then derive the Lie algebra of Poisson
brackets.

If we  seek  for a reduction of dynamics by cyclic coordinates,
the procedure is possible in the following way: $i)$  we
arbitrarily choose  one of the symmetries, or a linear combination
of them, searching for new coordinates where, as sketched above,
the cyclic variables appear. After the reduction, we get a new
Lagrangian $\widetilde{{\l}}({\bf Q})$; $ii)$ we search again for
symmetries in this new configuration space, make a new reduction
and so on until possible; $iii)$ if the search fails, we try again
by another of the existing symmetries.

Let us now assume that ${\cal L}$ is of the form (\ref{02}). As
${\bf X}$ is of the form (\ref{05}), $L_{\bf X}{\cal L}$ will be a
homogeneous polynomial of second degree in the velocities plus a
inhomogeneous term in the $q^{i}$. Since such a polynomial has to
be identically zero, each coefficient must be independently zero.
If $n$ is the dimension of the configuration space, we get
$\{1+n(n+1)/2\}$ partial differential equations. The system is
overdetermined, therefore, if any solution exists, it will be
expressed in terms of integration constants instead of boundary
conditions. It is also obvious that an overall constant factor in
the Lie vector ${\bf X}$ is irrelevant. In other words, the
Noether Symmetry Approach can be used to select functions which
assign the models and such functions (and then the models) can be
physically relevant.

Considering the specific case which we are going to discuss, the
$f(R)$ cosmology, the situation is the following. The
configuration space is ${\cal Q}=\{a,R\}$ while the tangent space
for the related tangent bundle is ${\cal
TQ}=\{a,\dot{a},R,\dot{R}\}$. The Lagrangian is an application
\beq {\cal L}: {\cal TQ}\longrightarrow \Re\eeq where $\Re$ is the
set of real numbers. The generator of symmetry is \beq
\label{vec}{\bf X}=\alpha\frac{\pa}{\pa a}+\beta\frac{\pa}{\pa
R}+\dot{\alpha}\frac{\pa}{\pa
\dot{a}}+\dot{\beta}\frac{\pa}{\pa\dot{R}}\,.\eeq As discussed
above,  a symmetry exists if the equation $L_{\bf X}{\cal L}=0$
has solutions. Then there will be a constant of motion on shell,
i.e.\ for the solutions of the Euler equations, as stated above
equation (\ref{09}).  In other words, a symmetry exists if at
least one of the functions $\alpha$ or $\beta$ in Eq.(\ref{vec})
is different from zero. As a byproduct,  the form of $f(R)$, not
specified in the point-like Lagrangian (\ref{eqz0}), is determined
in correspondence  to such a symmetry.

\section{Noether symmetries in $f(R)$ cosmology}\label{sec4}
For the existence of a symmetry, we can write the following system of equations (linear in $\alpha$ and $\beta$),
\bea
f_R\,(\a+2a\,\pa_a\a)+a\,f_{RR}\,(\b+a\,\pa_a\b)&=&0\label{eqz1}\\
a^2\,f_{RR}\,\pa_R\a&=&0\label{eqz2}\\
2\,f_R\,\pa_R\a+f_{RR}\,(2\,\a+a\,\pa_a\a+a\,\pa_R\b)+a\,\b\,f_{RRR}&=&0\label{eqz3}\,
, \eea obtained setting to zero the coefficients of the terms
$\dot{a}^2$, $\dot{R}^2$ and $\dot{a}\dot{R}$ in $L_{\bf X}{\cal
L}=0$.  In order to make $L_{\bf X}{\cal L}=0$ vanish we will also
look for those particular $f$'s which, given the Euler dynamics,
also satisfy the constraint \be
3\a\,(f-R\,f_R)-a\,\b\,R\,f_{RR}-\frac{6\kappa}{a^2}\,(\a
f_R+a\,\b\,f_{RR})\label{eqz4}+\frac{\rho_{r0}\,\a}{a^4}=0\, . \ee
This procedure is different from the usual Noether symmetry
approach, in the sense that now $L_{\bf X}{\cal L}=0$ will be
solved not for all dynamics (which solve the Euler-Lagrange
equations), but only for those $f$ which allows Euler solutions to
solve also the constraint (\ref{eqz4}). Imposing such a constraint
on the form of $f$ will turn out to be, as we will show, a sufficient condition to find solutions of the Euler-Lagrange
equation which also possess a constant of motion, i.e.\ a Noether
symmetry. As we shall see later on, the system (\ref{eqz1}),
(\ref{eqz2}) and (\ref{eqz3}) can be solved exactly. Having a
non-trivial solution for $\alpha$ and $\beta$ for this system, one
finds a constant of motion if also the constraint (\ref{eqz4}) is
satisfied. In fact, with these $\alpha$ and $\beta$, only those
Euler-Lagrange solutions which also satisfy equation (\ref{eqz4})
will have a constant of motion. However, this will not happen for
all $f(R)$'s. The task will be to find such forms of $f$.

A solution of (\ref{eqz1}), (\ref{eqz2}) and (\ref{eqz3})  exists
if explicit forms of $\alpha$, $\beta$ are found. If, at least one
of them is different from zero, a Noether symmetry exists.

If $f_{RR}\neq0$, Eq.(\ref{eqz2})  can be  immediately solved
being \be \a=\a(a)\, . \ee The case $f_{RR}=0$ is trivial since
corresponds to the standard GR. We can rewrite Eqs.(\ref{eqz1})
and (\ref{eqz3}) as follows \bea
f_R\left(\a+2a\,\frac{d\a}{da}\right)+a\,f_{RR}\,(\b+a\,\pa_a\b)&=&0\label{eqz1b}\\
f_{RR}\left(2\,\a+a\,\frac{d\a}{da}+a\,\pa_R\b\right)+a\,\b\,f_{RRR}&=&0\label{eqz3b}\,
. \eea Since $f=f(R)$, then $\partial f/\partial a=0$, even in the
case we consider $R=R(a)$,  it is possible to solve equation
(\ref{eqz3b}), by writing it as \be
\pa_R(\b\,f_{RR})=-f_{RR}\left(2\,\frac\a
a+\frac{d\a}{da}\right)\label{eqz3bbis}\, \ee whose general
solution can be written as \be
\b=-\left[\frac{2\a}a+\frac{d\a}{da}\right]\frac{f_R}{f_{RR}}+\frac{h(a)}{f_{RR}}\,
. \ee Therefore one finds that Eq.\ (\ref{eqz1b}) gives \be
f_R\left[\a-a^2\,\frac{d^2\a}{da^2}-a\,\frac{d\a}{da}\right]+a\,\left[h-a\frac{dh}{da}\right]=0\,
, \ee which has solution \be \a=c_1\,a+\frac{c_2}a\,\qquad{\rm
and}\qquad h=\frac{\bar c}a\, , \ee where, being $a$
dimensionless, $c_1$ and $c_2$ have the same dimensions. We can
further fix $\a$ to be dimensionless, this fixes the dimensions of
$\b$ to be $[\b]=M^2$. Then also $[\bar c]=M^2$, so that we have
\be
\b=-\left[3\,c_1+\frac{c_2}{a^2}\right]\frac{f_R}{f_{RR}}+\frac{\bar
c}{a\,f_{RR}}\, . \ee

We can now use the expressions for $\alpha$ and $\b$ into Eq.(\ref{eqz4}) as follows
\be
f_R=\frac{3\,a\,(c_1\,a^2+c_2)\,f-\bar c\,(a^2\,R+6 \kappa) }{2 a (c_2 R-6 c_1 \kappa
   )}+\frac{\left(c_1 a^2+c_2\right) \rho_{r0}}{2 a^4 (c_2 R-6 c_1 \kappa )}\, ,
%f_R=\frac{c_1\,a^2+c_2}{c_2\,R-6\,\kappa\,c_1}\,\frac{\rho_{r0}+3\,a^4\,f}{2\,a^4}\,
\label{fRab} \ee if $c_2\,R-6\,\kappa\,c_1\neq0$. It is clear
that, for a general $f$, it will not be possible to solve at the same time the Euler-Lagrange equation and this constraint. Therefore we have to use the Noether constraint in order to find the subset of those $f$ which make this possible. As
we shall see later, it is convenient to look for a parametric
solution in the form $\bigl[H(a),f\bigl(R(a)\bigr)\bigr]$. In this
case, since $f_R=f'/R'$, the Noether condition corresponds to the
following ODE \be
\frac{f'(a)}{R'(a)}=\frac{3\,a\,(c_1\,a^2+c_2)\,f(a)-\bar
c\,(a^2\,R(a)+6 \kappa) }{2 a (c_2\, R(a)-6 c_1 \kappa
   )}+\frac{\left(c_1 a^2+c_2\right) \rho_{r0}}{2 a^4 (c_2 R(a)-6 c_1 \kappa )}\, .
\label{NT1} \ee It should be noted that this change of variable is
defined only if $R'\neq0$,  that is if $R$ is not constant during
the evolution. When this happens Eq. (\ref{eqz4}) or (\ref{fRaF})
sets $a=a_0={\rm constant}$, which corresponds to an uninteresting
solution.

Any Euler-Lagrange solution, by definition, satisfies the Einstein equations.
However we will show that there are forms of $f(R)$, for which a
subset of those solution will also be a Noether solution. In fact,
Eq.(\ref{fRab}) can also be rewritten as \be
c_1\,a^2\,(\rho_{r0}+3\,a^4\,f+12\,\kappa\,a^2\,f_R)+c_2\,[\rho_{r0}+a^4\,(3\,f-2\,R\,f_R)]=\bar
c\,a^3\,(a^2\,R+6\kappa)\, . \label{fRaF} \ee Therefore we look for a family of solutions that, being a Noether symmetry, gives  a class
of  $f(R)$ models.

This symmetry implies the existence of the following constant of motion
\be
\a\,(6\,f_{RR}\,a^2\,\dot R+12\,f_R\,a\,\dot a)
+\b\,(6\,f_{RR}\,a^2\,\dot a)=6\,\mu^3_0={\rm constant},\label{abete} \ee
where $\mu_0$ has the dimensions of a mass. Equation (\ref{abete}) can be recast
in the form
\be \frac{d(f_R)}{dt}=f_{RR}\,\dot
R=\frac{\mu^3_0}{a\,(c_1\,a^2+c_2)}
+\frac{c_1\,a^2-c_2}{c_1\,a^2+c_2}\,f_R\,H
-\frac{\bar c\,a}{c_1\,a^2+c_2}\,H\, , \ee
or, using the time-parameter $a$
\be
a H(a) \left(\frac{f''(a)}{R'(a)}-\frac{f'(a)
   R''(a)}{R'(a)^2}\right)-\frac{\left(a^2
   c_1-c_2\right) H(a) f'(a)}{\left(c_1
   a^2+c_2\right) R'(a)}=\frac{\mu_0 ^3}{a \left(c_1
   a^2+c_2\right)}
-\frac{\bar c\,a}{c_1\,a^2+c_2}\,H(a)\, . \label{NT2} \ee Once
Eq.\ (\ref{NT1}) is solved, because  the Noether constraint is
satisfied, the solution $\bigl[H(a),f\bigl(R(a)\bigr)\bigr]$ will
automatically solve also (\ref{NT2}) for a particular $\mu_0$.
Equation (\ref{abete}) can be used to reduce the order of the
Friedmann equation. In fact, writing Eq.(\ref{energy}) as \be
f-6\,f_{RR}\,\dot R\,H
-6\,f_R\,H^2-f_R\left(R+\frac{6\kappa}{a^2}\right)-\frac{\rho_{m0}}{a^3}-\frac{\rho_{r0}}{a^4}=0\,
,\ee we have \be
 f-
\frac{12\,c_1\,a^2}{c_1\,a^2+c_2}\,f_R\,H^2\,
-f_R\left(R+\frac{6\kappa}{a^2}\right)
+\frac{6\,\bar c\,a}{c_1\,a^2+c_2}\,H^2=\frac{6\,\mu_0^3\,H}{a\,(c_1\,a^2+c_2)}+\frac{\rho_{m0}}{a^3}+\frac{\rho_{r0}}{a^4}\,
, \label{reduc} \ee
where $f_R$ is given by (\ref{fRab}).  We will use this relation in order to find
out exact cosmological solutions. Namely, we will search for
solutions depending on the constant of motion $\mu_0$ determined
by the  Noether symmetry.

\section{Exact cosmological solutions}\label{sec5}
In order to find out exact cosmological solutions, let us discuss
the Noether condition Eq.(\ref{fRaF}) and the dynamical system
(\ref{energy}),(\ref{frd1}) with respect to the
values of the integration constants $c_{1,2}$, the structural
parameters $k,\rho_{r0}, \rho_{m0}$ and the Noether charge
$\mu_0$. Beside cosmological solutions, also the explicit form of
$f(R)$ will result fixed in the various cases. As we shall see
later on, analytical solutions can be easily found for the case
when both $\bar c$ and $\mu_0$ vanish at the same time. Therefore
in all this section, except one subsection, we will set $\bar
c=0$.

\subsection{Case $c_1=0$}
In this case, the Noether condition (\ref{fRaF}) reduces to \be
2\,R\,f_R-3\,f=\frac{\rho_{r0}}{a^4}\, . \label{ntc10} \ee

\subsubsection{Vacuum and pure dust case}
In vacuum, or in the presence of  dust only, i.e.\ $\rho_{r0}=0$,
we find \be f=f_0\left(\frac{R}{R_0}\right)^{\!3/2}\, .
\label{fr32} \ee  This solution, for the vacuum case
$\rho_{r0}=\rho_{m0}=0$,  has been already found \cite{lambiase}.
The absence of a ghost imposes that $f_R<0$, i.e.\ $f_0>0$ since
$R_0<0$. In the case of dust and no radiation
($\rho_{m0}\neq0,\rho_{r0}=0$), one can substitute
Eq.(\ref{fr32}) into (\ref{reduc}) to find \be \left(\frac
R{R_0}\right)^{\!3/2} +\frac{18\kappa}{a^2\,R_0}\left(\frac
R{R_0}\right)^{\!1/2}
=-\frac{12\mu_0^3\,H}{c_2\,a\,f_0}-\frac{2\rho_{m0}}{a^3\,f_0}\, .
\label{frc10R} \ee
\begin{enumerate}
\item $k=0$. In this case, for consistency, we need the right hand side of (\ref{frc10R}) to be
 positive. If $\mu_0=0$ (case for which analytical solutions could be given), this is impossible as $f_0>0$, therefore there is no ghost-free solution. For the more general case $\mu_0^3/c_2<0$, there could be a physical solution: the non-linearity of the equations does not allow us to find analytical solutions for this case. Nevertheless, solutions (to be found numerically) may still exist.
\item $k\neq0$. The Ricci scalar can be found as the solution of Eq.\ (\ref{frc10R}). For $\mu_0=0$, we have a cubic equation in $(R/R_0)^{1/2}$, for which a real solution always exists (which may not be positive though). Looking at equation (\ref{frc10R}), the case $\mu_0=0,k=-1$ has no ghost-free solutions ($f_0<0$). Also the case $\mu_0=0,k=1$ has no solution, because we have
\be \sqrt{\frac R{R_0}}=\left[\frac{\tilde B_0^{1/3}}{f_0 R_0}-\frac{6\kappa\, f_0 }{\tilde B_0^{1/3}}\right]\frac1a\, , \ee
where we have defined the constant
\be \tilde B_0=\sqrt{f_0^4 \rho_{m0}^2 R_0^6 +216 f_0^6\, \kappa^3\,
R_0^3}-f_0^2 \rho_{m0} R_0^3\, , \ee
which implies that $(f_0/\rho_{m0})^2\,(\kappa/R_0)^3>-1/216$. If so, then, since $R_0<0$, $\tilde B_0>0$. However, this would lead to a negative value for $(R/R_0)^{1/2}$.
\end{enumerate}

\subsubsection{Dust and radiation case}

In this case we have  \be
f_R=\frac32\,\frac{f}{R}+\frac{\rho_{r0}}{2\,a^4\,R}\, .\label{reduc2bis} \ee
Once again, in order to have $f_R<0$, and $R<0$ during the evolution of the universe one requires
\be
f>-\frac{\rho_{r0}}{3\,a^4}\, .\label{constrN1}
\ee
If we substitute the expression for $f_R$ into the reduced Friedmann
Eq.(\ref{reduc}) we find
\be
f=-\frac{12\, \mu^3_0 \,a\,H\,R}{c_2
\left(R\, a^2+18
   \kappa\right)}-\frac{6 \kappa\, \rho_{r0}}{a^4 \left(R a^2+18 \kappa\right)}-\frac{3 \rho_{r0}
   R}{a^2 \left(R a^2+18 \kappa\right)}-\frac{2 \rho_{m0} R}{a \left(R a^2+18
   \kappa\right)}\, .
\label{fc10}
\ee
This relation gives $f$ as a function of $a$
being $R=R(a)$. It has to be $c_2\neq0$  otherwise the Noether
condition becomes trivial. This expression can be inserted back
into (\ref{reduc2bis}). Assuming $R=R(a)$ as a monotonic function of
$a$, one finds that $f_R=(df/da) / (dR/da)$, and equation
(\ref{ntc10}) becomes a differential equation for $R(a)$, which
can be written as \bea R'&=& \frac6{a^3 \left(18 a^3 H \mu_0 ^3+4
   c_2 \rho_{r0}+3 a c_2 \rho_{m0}\right) \left(R a^2+6 \kappa\right)}\times
\{-R^2 \,[2 a^3 \,(H-a H') \mu^3_0\notag\\
&+&c_2(2 \rho_{r0}+a \rho_{m0})] a^4 +6 \kappa R [6 a^3 \mu ^3_0\,(H+a
H')-c_2 (4 \rho_{r0}+a \rho_{m0})] a^2 -72 c_2 \kappa^2 \rho_{r0}\}\, ,
\label{Rpc10} \eea where the prime denotes differentiation with
respect to the scale factor $a$.  Eq.(\ref{Rpc10}) can be further
rewritten as a second order differential equation in $H(a)$, by
using equation (\ref{frd1}),
\be R=-12\,H^2-6\,a\,H\,H'-6\,\frac{\kappa}{a^2}\, . \label{Ra} \ee
Substituting (\ref{Ra}) into (\ref{Rpc10}) one finds \bea
H''&=&-\frac1{a^4 H^2 \left(18 a^3 H \mu_0 ^3+4 c_2 \rho_{r0}+3
   a c_2 \rho_{m0}\right)}\times\{24 a \kappa^2 \mu_0 ^3+H [a^2 \{a^2 (6 a^3 H
   \mu_0 ^3+4 c_2 \rho_{r0}\notag\\
&&{}+3 a c_2 \rho_{m0}) {H'}^2+a[12 a \kappa \mu_0 ^3+H (78 a^3 H \mu_0 ^3+32c_2 \rho_{r0}+21 a c_2 \rho_{m0})] H'\notag\\
&&{}+12 H[2 a \kappa \mu_0 ^3+H (2 a^3 H \mu_0 ^3+2 c_2\rho_{r0}+a c_2 \rho_{m0})]\}-8 c_2 \kappa\rho_{r0}]\}\, .
\label{CTc10}
\eea
 This differential equation selects those $f(R)$ models which satisfy, at the same
 time, both the Friedmann equation and the Noether condition. It has to be stressed that, having chosen $a$ as the time variable, finding the $H(a)$'s which solve (\ref{CTc10}) uniquely fixes the metric tensor. Hence, $H(a)$ represents a fully solved exact solution for the Einstein equations. Of course, if one wants to know the link between $a$ and the proper time, $a=a(t)$, one needs to find the integral $t=\int da/(aH)$.

The case $\mu_0=0$ is interesting as it allow us to find analytical solutions, as the differential equation becomes (2nd order and) linear for the variable $H^2$. In this case, the solution of the equation will be a
 family $H=H(a,d_1,d_2,c_2,\mu_0,\kappa,\rho_{r0},\rho_{m0})$,
 where $d_{1,2}$ are two constants coming from the integration of Eq.(\ref{CTc10}).
 In turn, by using Eq.(\ref{Ra}), it is possible to define a function $R=R(a,d_1,d_2,c_2,\mu_0,\rho_{r0},\rho_{m0})$,
 which can then be substituted into Eq.(\ref{fc10}) in order to find
 the explicit
 parametric form of $f(R)$, i.e. $f=f(a,d_1,d_2,c_2,\mu_0,\rho_{r0},\rho_{m0})$.
 In other words, we find the explicit parametric form for $f(R)$
 where the parameter used to describe the $f(R)$ is the scale factor $a$ (see also \cite{mimicking}
 for a comparison with observations. However,
 in that case, the adopted $f(R)$ models were
 constructed by phenomenological considerations and not derived from some first
principle, as the existence of symmetries as discussed here).

We can distinguish some relevant cases.
\begin{enumerate}
\item $k=0$, $\mu_0=0$. In this case, by exactly integrating equation (\ref{CTc10}),
we find \be
H^2=d_2\,\frac{d_1+8\,a\,\rho_{r0}+3\,\rho_{m0}\,a^2}{a^4}\, , \ee
where $d_{1,2}$ are integration constants, with $[d_1]=M^4$ and
$[d_2]=M^{-2}$. This expression for $H(a)$ together with (\ref{fc10}) and (\ref{Ra}) form a solution for the set of ODE's (\ref{FRDA}), and (\ref{NT1}), so that Eq.\ (\ref{NT2}) is satisfied giving $\mu_0=0$. Although this solution is analytical it cannot be accepted because it allows for a negative Newton constant. In fact, equation (\ref{constrN1}) cannot be satisfied by equation (\ref{fc10}) if $k=0,\mu_0=0$. However the non-linear case $\mu_0/c_2<0$ could actually lead to physical solutions (to be discussed elsewhere in a forthcoming paper). For the same reason, also the case $k=-1,\mu_0=0$ should be rejected.

\item $k=1$, $\mu_0=0$. As far as $R<-18\kappa/a^2$, the second term in the l.h.s.\ of equation (\ref{fc10}) becomes positive, allowing for the possibility of finding a physical solution. The integration of (\ref{CTc10}) leads to
\be
H^2=\left(\sqrt2\,d_1-\frac{32\,\rho_{r0}^2\,\kappa}{9\rho_{0m}^2}\right)\frac1{a^4}
+\left(8\,d_2\,\rho_{r0}-\frac{16\rho_{r0}\,\kappa}{3\rho_{m0}}\right)\frac1{a^3}
+\frac{3\,d_2\,\rho_{m0}}{a^2}\, ,
\ee
with $[d_1]=M^2$, and $[d_2]=M^{-2}$. In order to find $d_1$ and $d_2$ one can fit this formula with the standard Friedmann equation of GR with only matter, radiation and curvature. Therefore, one has to consider
\bea
\sqrt2\,d_1-\frac{32\,\rho_{r0}^2\,\kappa}{9\rho_{0m}^2}&=&H_0^2\,\Om_{r0}^{\rm eff}\, ,\\
8\,d_2\,\rho_{r0}-\frac{16\rho_{r0}\,\kappa}{3\rho_{m0}}&=&H_0^2\,\Om_{m0}^{\rm eff}\, ,\\
3\,d_2\,\rho_{m0}&=&H_0^2\,\Om_{k0}^{\rm eff}\, ,
\eea
but this system admits no solutions as one finds
\be
\kappa=\tfrac12\,H_0^2\,\Om_{k0}^{\rm eff}-\tfrac3{16}\,\frac{\rho_{m0}}{\rho_{r0}}\,H_0^2\,\Om_{m0}^{\rm eff}<0\,
\ee
using today's data \cite{Spergel:2006hy}.
\end{enumerate}

\subsection{Case $c_2=0$}
In this case, the Noether condition (\ref{fRaF}) reduces to
\be
\rho_{r0}+3\,a^4\,f+12\,\kappa\,a^2\,f_R=0\, . \label{ntc20}
\ee

\subsubsection{Vacuum and dust only case}

In this case we have $\rho_{r0}=0$, and a flat universe cannot be solution as one would obtain $f=0$. Considering $k\neq0$ one finds
\be
f_R=-\frac{a^2\,f}{4\,\kappa}\, .\label{frigo}
\ee
Since $f_R<0$ then $f$ is positive when $k<0$ and viceversa. Substituting this into the Friedmann equation one finds
\be
\{a^3 c_1 [(12 H^2+R)\, a^2+10
   \kappa ]\}\,f=4 \kappa\,(6 H \mu_0^3+c_1 \rho_{m0})\, .
\ee Restricting ourselves only to the study of the simple and
linear  case of a vanishing $\mu_0$, we can distinguish two cases
\begin{enumerate}
\item $\rho_{m0}=0,\mu_0=0$. In this case one needs to impose
\be R=-12H^2-10\,\frac\kappa{a^2}\, , \ee which, together with the
definition of $R$, gives \be H^2=2 d_1-\frac{2\kappa}{3a^2}\, ,
\ee where $d_1$ is a constant of integration with dimensions
$M^2$. This behavior describes a universe with only a cosmological
constant and curvature. Equation (\ref{ntc20}) can now be solved
for $f(a)$ giving \be
f=\frac{d_2}a=d_2\left[-\frac{R+24d_1}{2\kappa}\right]^{1/2}\, ,
\ee where $d_2$ is a constant of integration with dimensions
$M^4$.
\item $\rho_{m0}\neq0,\mu_0=0$.
In this case the Friedmann equation and (\ref{frigo}) give
\be
f=\frac{4 \kappa \, \rho_{m0}}{\left(12 H^2+R\right) a^5+10
   \kappa  a^3}\, .
\ee Substituting this expression in (\ref{frigo}), and using the
definition  for $R$ in terms of $H(a)$ one finds a linear 2nd
order differential equation in $H^2(a)$, which has solution \be
H^2=\frac{d_1}{2a^4}+2d_2-\frac{2\kappa}{3a^2}\, , \ee where
$d_{1,2}$ are integration constants, and $[d_1]=[d_2]=M^2$.
Therefore one has \bea
R&=&-24 d_2-\frac{2\kappa}{a^2}\,,\\
f&=&-\frac{2\kappa\,\rho_{m0}}{3\,a\,d_1}\, .
\eea
\end{enumerate}

\subsubsection{Radiation and dust case}
Also in this case, we have three possibilities, according to the
values of $k$.
\begin{enumerate}
\item $k=0.$ In this case one finds that \be f=-\frac{\rho_{r0}}{3
a^4}\, . \ee Therefore we have
\be
f_R=\frac{f'}{R'}=\tfrac43\,\frac{\rho_{r0}}{a^5\,R'}\, .
\ee
A well-behaved background evolution requires, with our conventions,
  $R'>0$, so that $f_R>0$. This means a negative effective
Newton constant, i.e.\ the solution cannot be accepted.

\item $k\neq0$. In this case, using equation (\ref{ntc20}) one finds
\be
f_R=-\frac{\rho_{r0}}{12\,\kappa\,a^2}-\frac{f\,a^2}{4\kappa}\, ,
\ee
and then using Friedmann equation (\ref{reduc}) one can solve for $f$, as follows
\be
f = \frac{-c_1 \left(12 H^2+R\right) \rho_{r0} a^2+12
   \kappa  \left(6 H \mu_0^3+c_1 \rho_{m0}\right) a+6 c_1 \kappa  \rho_{r0}}{3 a^4
  c_1 \left[\left(12 H^2+R\right) a^2+10 \kappa \right]}\, .
\ee By plugging  this relation into the Noether condition
(\ref{ntc20}),  and using the definition of $R$ in terms of $H,
H'$, and $a$,  one finds the following differential equation for
$H(a)$ \bea H''&=&\frac{a H \left[-\left(18 a H \mu_{0}^3+3 a c_1
   \rho_{m0}+4 c_1\rho_{r0}\right)
   {H'}^2 a^4-3 \left(a H \left(30 a H
   \mu_{0}^3+5 a c_1 \rho_{m0}+8 c_1
   \rho_{r0}\right)-4 \kappa  \mu_{0}^3\right)
   H'\, a^2\right.}{a^5 H^2
   \left(18 a H \mu_{0}^3+3 a c_1 \rho_{m0}+4 c_1 \rho_{r0}\right)}\notag\\
&&{}+\frac{\left.4 \kappa  \left(6 a H \mu_{0}^3+a
   c_1 \rho_{m0}+2 c_1 \rho_{r0}\right)\right]-8 \kappa ^2 \mu_{0}^3}{a^5 H^2
   \left(18 a H \mu_{0}^3+3 a c_1 \rho_{m0}+4 c_1 \rho_{r0}\right)}\, .
\label{eqzc20}
\eea
In the case $\mu_0=0,\rho_{m0}\neq0$, this differential equation can be exactly integrated to give
\be
H^2= \frac{256 \kappa  \rho_{r0}^3}{405 a^5 \rho_{m0}^3}+\frac{16 \kappa  \rho_{r0}^2}{27 a^4
   \rho_{m0}^2}+\frac{8 d_1 \rho_{r0}}{5
   a^5}-\frac{2 \kappa }{3 a^2}+\frac{3 \rho_{m0} d_1}{2
   a^4}+2 d_2\, ,
\ee where $d_{1,2}$ are two constants of integration with
dimensions $[d_1]=M^{-2}=[d_2]^{-1}$. It is interesting to note
the presence of a new cosmological term in this Friedmann
equation, which goes as $a^{-5}$, which would correspond to a
matter term with equation of state parameter $w=2/3$.

If $\mu_0=0,\rho_{m0}=0$, i.e.\ a universe filled  with radiation
only, equation (\ref{eqzc20}) has the following solution \be
H^2=2d_2+\frac{2d_1}{5a^5}-\frac{2\kappa}{3a^2}\, , \ee with
$[d_1]=[d_2]=M^2$.
\end{enumerate}

\subsection{Case $c_1,c_2\neq0$}

In this case, one can divide equation (\ref{fRaF}) by $c_1$
finding \be
f_R=\frac{a^2+c_3}{c_3\,R-6\,k}\,\frac{\rho_{r0}+3\,a^4\,f}{2\,a^4}\,
,\label{fRabis} \ee where $c_3=c_2/c_1\neq0$. This implies that
\be
f_{RR}\,\dot R=\frac{\tilde\mu^3_0}{a\,(a^2+c_3)}
+\frac{a^2-c_3}{a^2+c_3}\,f_R\,H\, ,
\ee
 where
$\tilde\mu_0^3=\mu_0^3/c_1$.

Friedmann equation Eq.(\ref{reduc}) can be rewritten as
\be
 f-
\frac{12\,a^2}{a^2+c_3}\,f_R\,H^2\,
-f_R\left(R+\frac{6k}{a^2}\right)=\frac{6\tilde\mu_0^3\,H}{a\,(a^2+c_3)}+\frac{\rho_{m0}}{a^3}+\frac{\rho_{r0}}{a^4}\,
. \label{reducbis}
\ee
By substituting (\ref{fRabis}) into
(\ref{reducbis}),  and solving for $f$, one finds
\bea f&=&
\frac{12\,\tilde \mu_0^3\, a^5\, H\, (6 k-c_3\, R)}{a^4
   \left(a^2+c_3\right) [3 \left(12 H^2+R\right)
   a^4+(30 k+c_3 R)\, a^2+18 c_3 k]}\notag\\
&&{}-\frac{\rho_{r0} \left(12 H^2+R\right) a^4+2 \rho_{m0}\, (c_3 R-6 k)\, a^3+3
   \rho_{r0}\, (c_3 R-2 k)\, a^2+6 c_3 k \rho_{r0}}{a^4\,
   [3 \left(12 H^2+R\right)
   a^4+(30 k+c_3 R)\, a^2+18 c_3 k]}\, ,
\label{genFa} \eea which means that the Noether symmetry, combined
with the dynamics,  determines  the form of $f$. In this case $f$
is a  function of $a$ since both $R$ and $H$ are functions of $a$.
We can still go further by using the same trick used  in the
previous section, i.e.\ considering $f$ as an implicit function of
$a$ into the Noether condition (\ref{fRabis}). Since $f=f(R(a))$
one finds \be
f_R=\frac{df}{dR}=\frac{da}{dR}\,\frac{df}{da}=\frac{f'}{R'}\, .
\label{dfdae} \ee Plugging Eqs.(\ref{genFa}) and (\ref{dfdae})
into (\ref{fRabis}), one finds a second order differential
equation for $H$, as follows \bea H''&=&\frac1{a^4
   (a^2+c_3)\,(3 a^2+c_3)\,
   H^2 \,[18\, \tilde\mu_0^3\,  H\, a^3+(a^2+c_3)\,
   (4 \rho_{r0}+3 a \rho_{m0})]}\notag\\
&&{}\times\left\{-24 c_3 \,(3 a^2+c_3)\, \tilde\mu_0^3\,  H^4
   \,a^5-24 \,(a^2+c_3)^2\, k^2\, \tilde\mu_0^3\,a\right.\notag\\
&&-H^2 \left[6 \,(3 a^2+c_3)^2\, \tilde\mu_0^3\,
   {H'}^2\, a^4
+24 \,(-3 a^4-2 c_3\,a^2+c_3^2)\, k\, \tilde\mu_0^3\right. \notag\\
&&\left.{}+(a^2+c_3)^2
   \,(45\, \rho_{m0}\, a^3+72 \,\rho_{r0}\, a^2+21\, c_3\, \rho_{m0}\, a+32\, c_3\,\rho_{r0})\, H'\right] a^3\notag\\
&&{}-6 H^3 \left[(3
   a^2+c_3) \,(15 a^2+13 c_3)\, \tilde\mu_0^3
   \,H'\, a^4+2 c_3 \,(a^2+c_3)^2\, (2 \rho_{r0}+a\rho_{m0})\right] a^2\notag\\
&&{}-(a^2+c_3)\, H\,\bigl[a^4\, H'\, [12
   (c_3-3 a^2)\, k\, \tilde\mu_0^3
   +(a^2+c_3)\, (3 a^2+c_3)
  \, (4 \rho_{r0}+3 a \rho_{m0})\, H']\notag\\
&&\left.{}-4 \,(a^2+c_3)\, k\,
   (3 \rho_{m0} a^3+6 \rho_{r0} a^2+2 c_3 \rho_{r0})\bigr]\right\}\, .
\label{FRDNT} \eea This differential equation defines the dynamics
of the Noether  solutions for a generic $f(R)$ model compatible
with the Noether symmetry. This result is  relevant since there is
a free parameter $c_3$, which together with the initial conditions
for $H_0$ and $H_0'$, uniquely specify the dynamics.  This
non-linear ODE is still of second order in $H(a)$ as the
$0,0$-Einstein equation for any $f(R)$ theory. However, there is a
huge improvement as this equation is independent of the explicit
form $f(R)$, having as the only unknown parameters two real
numbers, $c_3$ and $\mu_0$, the Noether charge. This also says
that for any value of the Noether charge there is a solution, the
solution of (\ref{FRDNT}). Therefore all the solutions of
(\ref{FRDNT}), as $c_3,\mu_0$ vary, represent the whole set of
Noether-charged cosmological solutions of the $f(R)$ theories.

\subsubsection{Vacuum and pure dust case}

In this case equation (\ref{fRabis}) reduces to
\be
f_R=\frac{3 f \left(a^2+c_3\right)}{2 \left(R\, c_3-6 \kappa \right)}\, ,
\label{relais1}
\ee
whereas $f$ can be written as
\be
f=\frac{2 \left(6 \kappa -R c_3\right) \left(\left(6 H \mu_{0}^3+\rho_{m0}\right) a^2+\rho_{m0}
   c_3\right)}{a \left(a^2+c_3\right) \left(3 \left(12
   H^2+R\right) a^4+\left(30 \kappa +R c_3\right) a^2+18 \kappa
   c_3\right)}\, .
\ee The case $\rho_{m0}=0,\mu_0=0$ admits no solutions, therefore,
as before, we will only discuss the case $\mu_0=0,\rho_{m0}\neq0$,
for which we can recast $f$ in the following form \be f=3 \left(12
H^2+R\right) a^4+\left(30 \kappa +R c_3\right) a^2+18
   \kappa  c_3\, .
\ee
Inserting this relation into (\ref{relais1}) together with the definition of $R$ one finds
\be
H''=\frac{-4 c_3 H^2-a \left(15 a^2+7 c_3\right) H'\,H
-a^2 \left(3 a^2+c_3\right) {H'}^2+4 \kappa }{a^2 \left(3 a^2+c_3\right)H}\, ,
\ee
whose general solution reads
\be
H^2=-\frac{c_3 \kappa }{9 a^4}-\frac{2 \kappa }{3 a^2}+\frac{2
   d_1}{a^4}+\frac{2 c_3 d_2}{a^2}+3 d_2\, .
\ee

\subsubsection{Pure radiation case}

Once again, studying Eq.\ (\ref{FRDNT}) to the case $\mu_0=0$  and
$\rho_{m0}=0$, we find the following equation \be
\bigl(H^2\bigr)''=-\frac{18 a^2+8 c_3}%
{a \left(3  a^2+c_3\right) }\,\bigl(H^2\bigr)'
-\frac{12\,c_3\, H^2}{a^2 \left(3  a^2+c_3\right)}
+\frac{2k(6  a^2+2c_3)}{a^4 \left(3  a^2+c_3\right)}\, .
\label{genc1c2mu0rhom0}
\ee
The general solution, when $c3>0$, for this ODE is
\bea
H^2&=&\frac{3 c_3 d_1}{a^4}+\frac{27
   d_1}{c_3}+\frac{18 d_1}{a^2}+\frac{5
   \sqrt{3} \sqrt{c_3} d_2}{a^3}+\frac{9 \sqrt{3}
   d_2}{a \sqrt{c_3}}+\frac{4 \kappa
   }{c_3}+\frac{2 \kappa }{a^2}\notag\\
&&{}+\frac{3 c_3
   d_2 \arctan\!\left(\frac{\sqrt{3}
   a}{\sqrt{c_3}}\right)}{a^4}+\frac{27 d_2 \arctan\!\left(\frac{\sqrt{3}
   a}{\sqrt{c_3}}\right)}{c_3}+\frac{18 d_2
   \arctan\!\left(\frac{\sqrt{3}
   a}{\sqrt{c_3}}\right)}{a^2}\, ,
\eea
whereas, for $c_3<0$, one finds
\bea
H^2&=&\frac{3 c_3 d_1}{a^4}+\frac{27
   d_1}{c_3}+\frac{18 d_1}{a^2}-\frac{5
   \sqrt{3} \sqrt{c_3} d_2}{a^3}+\frac{9 \sqrt{3}
   d_2}{a \sqrt{-c_3}}+\frac{4 \kappa
   }{c_3}+\frac{2 \kappa }{a^2}\notag\\
&&{}+\frac{3 c_3
   d_2 {\rm arctanh}\!\left(\frac{\sqrt{3}
   a}{\sqrt{-c_3}}\right)}{a^4}+\frac{27 d_2 {\rm arctanh}\!\left(\frac{\sqrt{3}
   a}{\sqrt{-c_3}}\right)}{c_3}+\frac{18 d_2
   {\rm arctanh}\!\left(\frac{\sqrt{3}
   a}{\sqrt{-c_3}}\right)}{a^2}\, .
\eea Either expression for $H(a)$ together with Eq.\ (\ref{genFa})
and  Eq.\ (\ref{Ra}) form a solution for (\ref{FRDA}), and
(\ref{NT1}), and possess $\mu_0=0$ Noether charge.

\subsubsection{Matter and Radiation case}

Let us restrict our study to the case $\tilde\mu=0$,  for which we
can find analytical solutions. Eq.(\ref{FRDNT}) reduces to \bea
\bigl(H^2\bigr)''&=&-\frac{\left(45 \rho_{m0} a^3+72  \rho_{r0} a^2+21 c_3 \rho_{m0} a+32 c_3 \rho_{r0}\right)}%
{a \left(3  a^2+c_3\right) (4 \rho_{r0}+3 a \rho_{m0})}\,\bigl(H^2\bigr)'\notag\\
&&{}-\frac{24\,c_3\left( \rho_{m0} a+2  \rho_{r0}
   \right) H^2}{a^2 \left(3  a^2+c_3\right) (4 \rho_{r0}+3 a \rho_{m0})}
+\frac{8k(3  \rho_{m0} a^3+6   \rho_{r0} a^2+2
   c_3 \rho_{r0})}{a^4 \left(3  a^2+c_3\right) (4 \rho_{r0}+3 a \rho_{m0})}\, .
\label{genc1c2mu0} \eea It is remarkable that this differential
equation is linear in  $H^2$. This makes the problem of solving it
much easier. In fact, analytical solutions for $k=0,\pm 1$ can be
achieved. Let us discuss them.
\begin{enumerate}
\item $k=0$. The solution of Eq.(\ref{genc1c2mu0}) is
\bea
H^2&=&\frac{4 d_1 d_2 c_3^{9/2}}{a^4}+\frac{24 d_1 d_2
   c_3^{7/2}}{a^2}-\frac{\rho_{0m} d_2
   c_3^{5/2}}{a^4}+36 d_1 d_2 c_3^{5/2}\notag\\
&&{}+\frac{2
   \sqrt{3} \rho_{r0} \arctan\!\left(\frac{\sqrt{3}
   a}{\sqrt{c_3}}\right) d_2
   c_3^2}{a^4}+\frac{10 \rho_{r0} d_2
   c_3^{3/2}}{a^3}
+\frac{12 \sqrt{3} \rho_{r0} \arctan\!\left(\frac{\sqrt{3} a}{\sqrt{c_3}}\right)
   d_2 c_3}{a^2}\notag\\
&&{}+\frac{18 \rho_{r0} d_2
   \sqrt{c_3}}{a}+18 \sqrt{3} \rho_{r0} \arctan\!\left(\frac{\sqrt{3} a}{\sqrt{c_3}}\right)
   d_2\, ,
\eea where $d_1$ and $d_2$ are integration constants with
dimensions, $[d_1]=M^4$, and $[d_2]=M^{-2}$. This is clearly a
deviation from standard GR, because there is a $1/a$ term, which
leads to an accelerated behavior if  dominates. Furthermore there
are terms, all involving $\rho_{r0}$, which include the arctangent
of $a$, where $c_3$ is supposed to be positive. These terms have
different behavior at low and high redshift. In fact since
${\displaystyle\lim_{a\to0}\arctan(a)\sim a}$ at high redshifts,
these terms behave as dust, $1/a$ and $a$ respectively, and are
subdominant with respect  to the radiation. On the other hand,
since  ${\displaystyle \lim_{a\to\infty}\arctan(a)\sim\pi/2}$ for
large and positive $a$, these terms will behave as radiation,
curvature and cosmological constant respectively.  It is also
interesting to notice that in order to have a true dust matter
component at late times, it has to be \be
10\,\rho_{r0}\,d_2\,c_3^{3/2}=\frac{8\pi G}3\,\rho_{m0}\,
.\label{dmk0} \ee This means that  $\rho_{r0}$ behaves as the
source of matter component in this modified Friedmann equation.  A
cosmological constant term is also present. It is determined by
the integration constants of the Noether condition.

As for the case $c_3<0$, the solution of Eq.\ (\ref{genc1c2mu0}) can be written as follows
\bea
H^2&=&-\frac{4 d_1 d_2 (-c_3)^{9/2}}{a^4}+\frac{24 d_1 d_2
   (-c_3)^{7/2}}{a^2}+\frac{\rho_{0m} d_2
   (-c_3)^{5/2}}{a^4}-36 d_1 d_2 (-c_3)^{5/2}\notag\\
&&{}+\frac{2
   \sqrt{3} \rho_{r0} {\rm arctanh}\!\left(\frac{\sqrt{3}
   a}{\sqrt{-c_3}}\right) d_2
   c_3^2}{a^4}+\frac{10 \rho_{r0} d_2
   (-c_3)^{3/2}}{a^3}
+\frac{12 \sqrt{3} \rho_{r0} {\rm arctanh}\!\left(\frac{\sqrt{3} a}{\sqrt{-c_3}}\right)
   d_2 c_3}{a^2}\notag\\
&&{}-\frac{18 \rho_{r0} d_2
   \sqrt{-c_3}}{a}+18 \sqrt{3} \rho_{r0} {\rm arctanh}\!\left(\frac{\sqrt{3} a}{\sqrt{-c_3}}\right)
   d_2\, .\label{solzH}
\eea For this solution, as a pedagogical example,  more detailed
calculations and a link with scalar-tensor theories are given in
the appendix.

\item $k\neq0$. The general solution is
\bea
H^2&=&-\frac{32 \kappa  \arctan\left(\frac{\sqrt{3} a}{\sqrt{c_3}}\right)\rho_{r0}^3}{9 \sqrt{3} a^4 \rho_{m0}^3
   \sqrt{c_3}}-\frac{160 \kappa  \rho_{r0}^3}{27 a^3 \rho_{m0}^3 c_3}-\frac{64 \kappa  \arctan\left(\frac{\sqrt{3}
   a}{\sqrt{c_3}}\right) \rho_{r0}^3}{3 \sqrt{3} a^2 \rho_{m0}^3 c_3^{3/2}}-\frac{32 \kappa  \rho_{r0}^3}{3 a \rho_{m0}^3 c_3^2}-\frac{32 \kappa  \arctan\left(\frac{\sqrt{3} a}{\sqrt{c_3}}\right) \rho_{r0}^3}{\sqrt{3} \rho_{m0}^3
   c_3^{5/2}}\notag\\
&&{}-\frac{16 \kappa  \rho_{r0}^2}{3 a^2 \rho_{m0}^2 c_3}-\frac{8 \kappa  \rho_{r0}^2}{27 a^4 \rho_{m0}^2}
-\frac{8 \kappa  \rho_{r0}^2}{\rho_{m0}^2 c_3^2}+\frac{\sqrt{3} \arctan\left(\frac{\sqrt{3} a}{\sqrt{c_3}}\right)
   d_2 \rho_{r0}}{2 a^4 c_3^{5/2}}+\frac{5 d_2 \rho_{r0}}{2 a^3 c_3^3}+\frac{3 \sqrt{3} \arctan\left(\frac{\sqrt{3}
   a}{\sqrt{c_3}}\right) d_2 \rho_{r0}}{a^2 c_3^{7/2}}\notag\\
&&{}+\frac{9 d_2 \rho_{r0}}{2 a c_3^4}+\frac{9 \sqrt{3} \arctan\left(\frac{\sqrt{3} a}{\sqrt{c_3}}\right) d_2 \rho_{r0}}{2 c_3^{9/2}}-\frac{2 \kappa  \arctan\left(\frac{\sqrt{3}
   a}{\sqrt{c_3}}\right) \sqrt{c_3} \rho_{r0}}{\sqrt{3} a^4 \rho_{m0}}-\frac{4 \sqrt{3} \kappa  \arctan\left(\frac{\sqrt{3}
   a}{\sqrt{c_3}}\right) \rho_{r0}}{a^2 \rho_{m0} \sqrt{c_3}}\notag\\
&&{}-\frac{10 \kappa  \rho_{r0}}{3 a^3 \rho_{m0}}-\frac{6
   \kappa  \rho_{r0}}{a \rho_{m0} c_3}-\frac{6 \sqrt{3} \kappa  \arctan\left(\frac{\sqrt{3} a}{\sqrt{c_3}}\right) \rho_{r0}}{\rho_{m0} c_3^{3/2}}-\frac{2 \kappa }{3 a^2}-\frac{\kappa  c_3}{9 a^4}+\frac{6 d_1}{a^2 c_3}+\frac{9
   d_1}{c_3^2}+\frac{d_1}{a^4}-\frac{\rho_{m0} d_2}{4 a^4 c_3^2}\, .
\eea
%\bea
%H^2&=&\frac{8 d_1 c_3^4}{a^4}+\frac{48 d_1 c_3^3}{a^2}+\frac{72 d_1 c_3^3}{a^4}
%-\frac{\rho_{m0} d_2 c_3^{5/2}}{a^4}+\frac{432 d_1
%   c_3^2}{a^2}+\frac{144 d_1 c_3^2}{a^4}+72 d_1 c_3^2\notag\\
%&&{}+\frac{2 \sqrt{3} \rho_{r0} \arctan\!\left(\frac{\sqrt{3}
%   a}{\sqrt{c_3}}\right) d_2 c_3^2}{a^4}+\frac{10 \rho_{r0} d_2 c_3^{3/2}}{a^3}-\frac{4 c_3}{9 a^4}+\frac{864 d_1
%   c_3}{a^2}+648 d_1 c_3\notag\\
%&&{}+\frac{12 \sqrt{3} \rho_{r0} \arctan\!\left(\frac{\sqrt{3} a}{\sqrt{c_3}}\right) d_2 c_3}{a^2}+\frac{18
%   \rho_{r0} d_2 \sqrt{c_3}}{a}-\frac{2}{3 a^2}+1296 d_1\notag\\
%&&{}+18 \sqrt{3} \rho_{r0} \arctan\!\left(\frac{\sqrt{3} a}{\sqrt{c_3}}\right) d_2-\frac{8
%   \rho_{r0}^2}{9 a^4 \rho_{m0}^2}-\frac{16 \rho_{r0}^2}{3 a^2 \rho_{m0}^2 c_3}-\frac{8 \rho_{r0}^2}{\rho_{m0}^2 c_3^2}\, ,
%\eea
Also in these cases we have interesting behaviors matching the
main cosmological eras. The integration constants $d_{1,2}$ have
dimensions respectively $[d_1]=M^2$, and $[d_2]=M^{-2}$.  The
analysis, for both this and the previous case ($k=0$), of the set
of parameters $\{d_1,d_2, c_3\}$ which can be bounded by
observations will be done in a forthcoming paper.
%\item $k=-1$. In this case, the general solution is
%%\bea
%H^2&=&\frac{8 d_1 c_3^4}{a^4}+\frac{48 d_1 c_3^3}{a^2}+\frac{72 d_1 c_3^3}{a^4}-\frac{\rho_{m0} d_2 c_3^{5/2}}{a^4}+\frac{432 d_1
%   c_3^2}{a^2}+\frac{144 d_1 c_3^2}{a^4}+72 d_1 c_3^2\notag\\
%&&{}+\frac{2 \sqrt{3} \rho_{r0} \arctan\!\left(\frac{\sqrt{3}
%   a}{\sqrt{c_3}}\right) d_2 c_3^2}{a^4}+\frac{10 \rho_{r0} d_2 c_3^{3/2}}{a^3}+\frac{4 c_3}{9 a^4}+\frac{864 d_1
%   c_3}{a^2}+648 d_1 c_3\notag\\
%&&{}+\frac{12 \sqrt{3} \rho_{r0} \arctan\!\left(\frac{\sqrt{3} a}{\sqrt{c_3}}\right) d_2 c_3}{a^2}+\frac{18
%   \rho_{r0} d_2 \sqrt{c_3}}{a}+\frac{2}{3 a^2}+1296 d_1\notag\\
%&&{}+18 \sqrt{3} \rho_{r0} \arctan\!\left(\frac{\sqrt{3} a}{\sqrt{c_3}}\right) d_2+\frac{8
%   \rho_{r0}^2}{9 a^4 \rho_{m0}^2}+\frac{16 \rho_{r0}^2}{3 a^2 \rho_{m0}^2 c_3}+\frac{8 \rho_{r0}^2}{\rho_{m0}^2 c_3^2}\, .
%\eea

Eq.\ (\ref{genc1c2mu0}), for the case $c_3<0$, has solution
\bea
H^2&=&\frac{32 \kappa  {\rm arctanh}\left(\frac{\sqrt{3} a}{\sqrt{-c_3}}\right)\rho_{r0}^3}{9 \sqrt{3} a^4 \rho_{m0}^3
   \sqrt{-c_3}}-\frac{160 \kappa  \rho_{r0}^3}{27 a^3 \rho_{m0}^3 c_3}-\frac{64 \kappa  {\rm arctanh}\left(\frac{\sqrt{3}
   a}{\sqrt{-c_3}}\right) \rho_{r0}^3}{3 \sqrt{3} a^2 \rho_{m0}^3 (-c_3)^{3/2}}-\frac{32 \kappa  \rho_{r0}^3}{3 a \rho_{m0}^3 c_3^2}+\frac{32 \kappa  {\rm arctanh}\left(\frac{\sqrt{3} a}{\sqrt{-c_3}}\right) \rho_{r0}^3}{\sqrt{3} \rho_{m0}^3
   (-c_3)^{5/2}}\notag\\
&&{}-\frac{16 \kappa  \rho_{r0}^2}{3 a^2 \rho_{m0}^2 c_3}-\frac{8 \kappa  \rho_{r0}^2}{27 a^4 \rho_{m0}^2}
-\frac{8 \kappa  \rho_{r0}^2}{\rho_{m0}^2 c_3^2}-\frac{\sqrt{3} {\rm arctanh}\left(\frac{\sqrt{3} a}{\sqrt{-c_3}}\right)
   d_2 \rho_{r0}}{2 a^4 (-c_3)^{5/2}}+\frac{5 d_2 \rho_{r0}}{2 a^3 c_3^3}+\frac{3 \sqrt{3} {\rm arctanh}\left(\frac{\sqrt{3}
   a}{\sqrt{-c_3}}\right) d_2 \rho_{r0}}{a^2 (-c_3)^{7/2}}\notag\\
&&{}+\frac{9 d_2 \rho_{r0}}{2 a c_3^4}-\frac{9 \sqrt{3} {\rm arctanh}\left(\frac{\sqrt{3} a}{\sqrt{-c_3}}\right) d_2 \rho_{r0}}{2 (-c_3)^{9/2}}-\frac{2 \kappa  {\rm arctanh}\left(\frac{\sqrt{3}
   a}{\sqrt{-c_3}}\right) \sqrt{-c_3} \rho_{r0}}{\sqrt{3} a^4 \rho_{m0}}+\frac{4 \sqrt{3} \kappa  {\rm arctanh}\left(\frac{\sqrt{3}
   a}{\sqrt{-c_3}}\right) \rho_{r0}}{a^2 \rho_{m0} \sqrt{-c_3}}\notag\\
&&{}-\frac{10 \kappa  \rho_{r0}}{3 a^3 \rho_{m0}}-\frac{6
   \kappa  \rho_{r0}}{a \rho_{m0} c_3}-\frac{6 \sqrt{3} \kappa  {\rm arctanh}\left(\frac{\sqrt{3} a}{\sqrt{-c_3}}\right) \rho_{r0}}{\rho_{m0} (-c_3)^{3/2}}-\frac{2 \kappa }{3 a^2}-\frac{\kappa  c_3}{9 a^4}+\frac{6 d_1}{a^2 c_3}+\frac{9
   d_1}{c_3^2}+\frac{d_1}{a^4}-\frac{\rho_{m0} d_2}{4 a^4 c_3^2}\, .
\eea
\end{enumerate}
It is worthy to note that once the free parameters are constrained
by the data (the set of allowed parameters might be empty anyhow),
 one can select physically interesting $f(R)$ models
as in \cite{mimicking}.

\subsubsection{Non-linear case, $\tilde\mu_0\neq0$}

In this more general case, Eq.(\ref{FRDNT}) cannot be written as a
linear differential equation in $H^2$, therefore it is not
possible to achieve an analytical general solution. However,
after fixing initial conditions for $H$ and giving suitable values for
the parameters, one can solve it numerically. These initial
conditions fix, in turn, the $f(R)$ model and  the behavior of
$H(a)$.

\subsubsection{General non-linear case, $\bar c\neq0$ and $\tilde\mu_0\neq0$}

By using Eq. (\ref{fRab}) inside Eq.\ (\ref{reduc}) one finds the following expression for $f$
\bea
f&=&\frac{c_1 \bar c R \left(12 H^2+R\right) a^5}{\left(c_1 a^2+c_2\right) \Delta}+\frac{\bar c
   R \left(12 c_2 H^2+12 c_1 \kappa +c_2 R\right) a^3}{\left(c_1 a^2+c_2\right) \Delta}\notag\\
&&{}+\frac{2 \left(36 c_1 \kappa  H \mu_0 ^3-6 c_2 H R \mu_0 ^3+18 c_1 \bar c \kappa ^2+6 c_1^2
   \kappa  \rho_{m0}+6 c_2 \bar c \kappa  R-c_1 c_2 \rho_{m0} R\right) a}{\left(c_1
   a^2+c_2\right) \Delta}\notag\\
&&{}-\frac{2 c_2 \left(-18 \bar c \kappa ^2-6 c_1 \rho_{m0} \kappa +c_2
   \rho_{m0} R\right)}{\left(c_1 a^2+c_2\right) \Delta\, a}-\frac{\rho_{r0} \left(12 c_1
   H^2 a^4+c_1 R a^4-6 c_1 \kappa  a^2+3 c_2 R a^2+6 c_2 \kappa \right)}{\Delta\, a^4}\, ,
\eea where \be \Delta=36 c_1 H^2 a^4+3 c_1 R a^4+30 c_1 \kappa
a^2+c_2 R a^2+18 c_2 \kappa\, . \ee The Friedmann equation gives us
the expression of $f$ in terms of $R(a)$, $H(a)$  and $a$. Eq.\
(\ref{NT1}), which can be rewritten here as \be
\frac{f'(a)}{R'(a)}=\frac{3\,a\,(c_1\,a^2+c_2)\,f(a)-\bar
c\,(a^2\,R(a)+6 \kappa) }{2 a (c_2\, R(a)-6 c_1 \kappa
   )}+\frac{\left(c_1 a^2+c_2\right) \rho_{r0}}{2 a^4 (c_2 R(a)-6 c_1 \kappa )}\, ,
\ee
giving a dynamics for $f$, defines a second order differential equation for $H$, given by
\bea
H''&=&\left[\left(c_1 a^2+c_2\right) H \Gamma\right]^{\!-1} {H'}^2 \left(12 c_1^2 {\bar c} \kappa  a^7+9 c_1^3 \rho_{m0} a^7+54 c_1^2 \mu ^3 H a^7+12 c_1^3
   \rho_{r0} a^6+24 c_1 c_2 {\bar c} \kappa  a^5+21 c_1^2 c_2 \rho_{m0} a^5\right.\notag\\
&&\left.{}+36 c_1
   c_2 \mu ^3 H a^5+28 c_1^2 c_2 \rho_{r0} a^4+12 c_2^2 {\bar c} \kappa  a^3+15 c_1
   c_2^2 \rho_{m0} a^3+6 c_2^2 \mu ^3 H a^3+20 c_1 c_2^2 \rho_{r0} a^2+3 c_2^3
   \rho_{m0} a+4 c_2^3 \rho_{r0}\right) \notag\\
&&{}  -\left[a
   \left(c_1 a^2+c_2\right) H \Gamma\right]^{\!-1}{H'}\left(54 c_1^2 {\bar c} H^3 a^9+108 c_1 c_2 {\bar c} H^3 a^7-270 c_1^2 \mu ^3 H^2
   a^7-60 c_1^2 {\bar c} \kappa  H a^7-45 c_1^3 \rho_{m0} H a^7\right.\notag\\
&&\left.{}-72 c_1^3 \rho_{r0} H
   a^6+36 c_1^2 \kappa  \mu ^3 a^5+54 c_2^2 {\bar c} H^3 a^5-324 c_1 c_2 \mu ^3 H^2 a^5-120
   c_1 c_2 {\bar c} \kappa  H a^5-111 c_1^2 c_2 \rho_{m0} H a^5\right.\notag\\
&&\left.{}-176 c_1^2 c_2
   \rho_{r0} H a^4+24 c_1 c_2 \kappa  \mu ^3 a^3-78 c_2^2 \mu ^3 H^2 a^3-60 c_2^2 {\bar c}
   \kappa  H a^3-87 c_1 c_2^2 \rho_{m0} H a^3-136 c_1 c_2^2 \rho_{r0} H a^2\right.\notag\\
&&\left.{}-12
   c_2^2 \kappa  \mu ^3 a-21 c_2^3 \rho_{m0} H a-32 c_2^3 \rho_{r0} H\right)
-\left[a^4 \left(c_1
   a^2+c_2\right) H^2 \Gamma\right]^{\!-1}4 \left(-18 c_1 c_2 \mu ^3 H^4 a^7-3 c_1^2
   c_2 \rho_{m0} H^3 a^7\right.\notag\\
&&\left.{}+18 c_1^2 \kappa  \mu ^3 H^2 a^7+6 c_1^2 {\bar c} \kappa ^2 H a^7+3
   c_1^3 \kappa  \rho_{m0} H a^7-6 c_1^2 c_2 \rho_{r0} H^3 a^6+6 c_1^3 \kappa
   \rho_{r0} H a^6-6 c_2^2 \mu ^3 H^4 a^5-6 c_1^2 \kappa ^2 \mu ^3 a^5\right.\notag\\
&&\left.{}-6 c_1 c_2^2
   \rho_{m0} H^3 a^5+12 c_1 c_2 \kappa  \mu ^3 H^2 a^5+12 c_1 c_2 {\bar c} \kappa ^2 H
   a^5+6 c_1^2 c_2 \kappa  \rho_{m0} H a^5-12 c_1 c_2^2 \rho_{r0} H^3 a^4+14
   c_1^2 c_2 \kappa  \rho_{r0} H a^4\right.\notag\\
&&\left.{}-12 c_1 c_2 \kappa ^2 \mu ^3 a^3-3 c_2^3 \text{$\rho
   $0m} H^3 a^3-6 c_2^2 \kappa  \mu ^3 H^2 a^3+6 c_2^2 {\bar c} \kappa ^2 H a^3+3 c_1 c_2^2
   \kappa  \rho_{m0} H a^3-6 c_2^3 \rho_{r0} H^3 a^2\right.\notag\\
&&\left.{}+10 c_1 c_2^2 \kappa  \rho_{r0}
   H a^2-6 c_2^2 \kappa ^2 \mu ^3 a+2 c_2^3 \kappa  \rho_{r0} H\right)\, ,
\eea
where
\bea
\Gamma&=&18 c_1 \bar c H^2 a^7+18 c_2 \bar c H^2 a^5-12 c_1 \bar c \kappa  a^5-9 c_1^2 \rho_{m0} a^5
-54 c_1 \mu ^3 H a^5-12 c_1^2 \rho_{r0} a^4-12 c_2 \bar c \kappa  a^3-12 c_1
   c_2 \rho_{m0} a^3\notag\\
&&{}-18 c_2 \mu ^3 H a^3-16 c_1 c_2 \rho_{r0} a^2-3 c_2^2
   \rho_{m0} a-4 c_2^2 \rho_{r0}\, .
\eea It is evident that a more detailed (numerical) study,
pursued elsewhere, of this differential equation is necessary in
order to study the dynamics of these solutions.

\subsection{Non-Noether solutions}

In general it is not possible  to find a solution of the Friedmann
equations which is also a Noether symmetry since, in principle,
such symmetries do not exist for any $f(R)$ theory. In general, a
solutions of the cosmological equations is not a solution
compatible with the condition $L_{{{\bf X}}}{\cal L}=0$. This is a
peculiar situation which holds only if conserved quantities
(Noether's charges) are intrinsically present in the structure of
the theory (in our case, the form of $f(R)$). For example,
imposing a power law solution, $a\propto t^p$, defines a function
of $R=R(a)$, which can be put in the Noether symmetry equations,
in order to find $f=f(R(a))$. Finally one can substitute the
expressions for $f(a)$, $R(a)$, and $H$ in the Friedmann
equations. In doing this, it is easy to show that, for $k=0$,
there are no simple power-law solutions compatible with a Noether
charge.

The method discussed above allows to  discriminate theories which
admit or not cosmological solutions compatible  with a Noether
charge.

It is also clear that power-law solutions do  exist in general for
$f(R)$ models, but they can be found using  different methods
\cite{noi}. Assuming, in general, a power-law $H(a)$, one finds
$R$ as a function of $a$, and then, in principle, $f=f(R(a))$. It
is therefore possible to write the Einstein equation as a second
order differential equation for $f$ as a function of $a$, whereas
all other quantities ($H$ and $R$) are given functions of $a$. The
same argument holds for the redshift $z$ \cite{mimicking}.

For example, let us rewrite the Friedmann equation (\ref{energy})
as  \be
 f-6\,f_{RR}\,\dot R\,H
-6\,f_R\,H^2-f_R\left(R+\frac{6k}{a^2}\right)=\frac
{\rho_{m0}}{a^3}+\frac{\rho_{r0}}{a^4}\, , \ee and let us consider
$H=\bar H(a)$ and $R=\bar R(a)$ as  given functions of $a$, being,
as above, \be \bar R=-12\,\bar H^2-6\,a\,\bar H\,\bar
H'-6\,\frac{k}{a^2}\, . \label{Rabis} \ee
 The Friedmann equation can be written as
\be f'' +\left[ \frac1a -\frac{\bar R''}{\bar R'} +\frac1{6a\,\bar
H^2}\left(\bar R+\frac{6k}{a^2}\right) \right]f' -\frac{\bar
R'}{6a\,\bar H^2}\,f=-\frac{\rho_{m0}\,a+\rho_{r0}}{6\,a^5\,\bar
H^2}\,\bar R'\, . \ee This is a  second order linear equation in
$f$, whose general solutions depends on two parameters, $f_0$ and
$f'_0$. Specifically, being the equation linear, the general
solution is the linear combination of two solutions of the
homogeneous ODE plus a particular solution. It is then clear that
more than one $f(R)$ model can have the same behavior for $H(a)$,
i.e.\ more theories share the same cosmological evolution.  This
situation is due to the fact that  one has a fourth-order gravity
theory. The singular points of this differential equation are
those for which either $\bar H$ or $d\bar R/da$ vanishes.

Starting from these considerations,  interesting classes of
solutions can be found out.

\subsubsection{Radiation solutions}

Let us seek for all the $f(R)$ models which have the particular
solution $a=\sqrt{t/t_0}$, which means \be \bar
H=\frac1{2\,t_0\,a^2}=\frac{H_0}{a^2}\, ,\qquad\textrm{so
that}\qquad \bar R=-\frac{6k}{a^2}\, , \ee where
$H_0\equiv(2\,t_0)^{-1}$. We have three interesting cases.
\begin{enumerate}
\item For $k=0$, we have $R=0$,
leading to the  Friedmann equation \be f(0)-6\,f_R(0)\,\bar
H^2=\frac{\rho_{m0}}{a^3}+\frac{\rho_{r0}}{a^4}\, , \ee which, if
$\rho_{m0}\neq0$, cannot be solved for $\bar H\sim a^{-2}$ since
$f(0)$ and $f'(0)$ cannot be functions of $a$, but only constants.
If $\rho_{m0}=0$, standard GR is of course recovered.
\item For the case $k=-1$ we have the following differential equation for $f$,
\be
f''+\frac{4}{a}\,f'+\frac{2\,\kappa}{H_0^2}\, f=\frac{2\,\kappa\, (\rho_{r0}+a \rho_{m0})}{H_0^2\,a^4}\, ,
\ee
whose general solution can be written as
\bea
R&=&-\frac{6\,\kappa}{a^2}\\
f&=&
\frac{\sqrt{\frac{a \sqrt{-\kappa}}{H_0}} d_2 \cos\!
   \left(\frac{a \sqrt{-2\kappa}}{H_0}\right)
   H_0^2}{\sqrt[4]{2} a^{7/2} \kappa \sqrt{\pi
   }}
-\frac{\sqrt{\frac{a \sqrt{-\kappa}}{H_0}} d_1 \sin\!
   \left(\frac{a \sqrt{-2\kappa}}{H_0}\right)
   H_0^2}{\sqrt[4]{2} a^{7/2} \kappa \sqrt{\pi
   }}
-\frac{\sqrt[4]{2} \sqrt{\frac{a \sqrt{-\kappa}}{H_0}} d_1 \cos\!
   \left(\frac{a \sqrt{-2\kappa}}{H_0}\right)
   H_0}{a^{5/2} \sqrt{-\kappa} \sqrt{\pi }}\notag\\
&&{}-\frac{\sqrt[4]{2}
   \sqrt{\frac{a \sqrt{-\kappa}}{H_0}} d_2 \sin\! \left(\frac{a
   \sqrt{-2\kappa}}{H_0}\right) H_0}{a^{5/2} \sqrt{-\kappa} \sqrt{\pi
   }}
+\frac{\rho_{m0}}{a^3}+\frac{\rho_{r0} \sqrt{-\kappa} \text{Ci}\!\left(\frac{\sqrt{2} a
   \sqrt{-\kappa}}{H_0}\right) \sin\! \left(\frac{a
   \sqrt{-2\kappa}}{H_0}\right)}{\sqrt{2} a^3 H_0}\notag\\
&&{}-\frac{\rho_{r0} \sqrt{-\kappa} \cos\!
   \left(\frac{ a \sqrt{-2\kappa}}{H_0}\right)
   \text{Si}\!\left(\frac{a
   \sqrt{-2\kappa}}{H_0}\right)}{\sqrt{2} a^3 H_0}
+\frac{\rho_{r0} \kappa
   \cos\!\left(\frac{a \sqrt{-2\kappa}}{H_0}\right)
   \text{Ci}\!\left(\frac{a \sqrt{-2\kappa}}{H_0}\right)}{a^2
   H_0^2}\notag\\
&&{}+\frac{\rho_{r0} \kappa \sin\! \left(\frac{a
   \sqrt{-2\kappa}}{H_0}\right) \text{Si}\!\left(\frac{a
   \sqrt{-2\kappa}}{H_0}\right)}{a^2 H_0^2}\, ,
\eea
where the SinIntegral and CosIntegral functions, Si and Ci respectively, are defined as
\be
{\rm Si}(x)=\int_0^x \frac{\sin(t)}t\,dt\,\qquad
{\rm Ci}(x)=-\int_x^\infty\frac{\cos(t)}{t}\, dt\, .
\ee
The integration constants $d_{1,2}$ have dimensions $[d_1]=[d_2]=M^4$.
\item Along the same lines, the case $k=1$ has the following solution
\bea
R&=&-\frac{6\,\kappa}{a^2}\\
f&=&\frac{\sqrt{\frac{a \sqrt{\kappa}}{H_0}} d_1 \cosh\!
   \left(\frac{\sqrt{2 \kappa} a }{H_0}\right)
   H_0^2}{\sqrt[4]{2} a^{7/2} \kappa \sqrt{\pi
   }}
+\frac{\sqrt{\frac{a \sqrt{\kappa}}{H_0}} d_1 \sinh\!
   \left(\frac{\sqrt{2 \kappa} a }{H_0}\right)
   H_0^2}{\sqrt[4]{2} a^{7/2} \kappa \sqrt{\pi
   }}
-\frac{\sqrt[4]{2} \sqrt{\frac{a \sqrt{\kappa}}{H_0}} d_1 \cosh\!
   \left(\frac{\sqrt{2 \kappa} a}{H_0}\right)
   H_0}{a^{5/2} \sqrt{\pi \kappa }}\notag\\
&&{}-\frac{\sqrt[4]{2}
   \sqrt{\frac{a \sqrt{\kappa}}{H_0}} d_1 \sinh\! \left(\frac{\sqrt{2 \kappa} a
   }{H_0}\right) H_0}{a^{5/2} \sqrt{\pi \kappa
   }}
+\frac{\rho_{m0}}{a^3}
-\frac{\rho_{r0} \sqrt{\kappa}\, \text{Chi}\!\left(\frac{\sqrt{2 \kappa} a
   }{H_0}\right) \sinh\! \left(\frac{\sqrt{2 \kappa} a
   }{H_0}\right)}{\sqrt{2} a^3 H_0}\notag\\
&&{}+\frac{\rho_{r0} \sqrt{\kappa} \cosh\!
   \left(\frac{\sqrt{2 \kappa} a}{H_0}\right)
   \text{Shi}\!\left(\frac{\sqrt{2 \kappa} a}{H_0}\right)}{\sqrt{2} a^3 H_0}
-\frac{\rho_{r0} \kappa
   \cosh\! \left(\frac{\sqrt{2 \kappa} a}{H_0}\right)
   \text{Chi}\!\left(\frac{\sqrt{2 \kappa} a}{H_0}\right)}{a^2
   H_0^2}\notag\\
&&{}+\frac{\rho_{r0} \kappa \sinh\! \left(\frac{\sqrt{2 \kappa} a}{H_0}\right) \text{Shi}\!\left(\frac{\sqrt{2 \kappa} a}{H_0}\right)}{a^2 H_0^2}\, ,
\eea
where the hyperbolic SinIntegral and CosIntegral, Shi and Chi
respectively, are defined as \be {\rm Shi}(x)=\int_0^x
\frac{\sinh(t)}t\,dt\,\qquad {\rm
Chi}(x)=\gamma_{E,M}+\ln(x)+\int_0^x\frac{\cosh(t)-1}{t}\, dt\, ,
\ee and $\gamma_{E,M}\approx0.577$ is the Euler-Mascheroni
constant.  Both $d_1$ and $d_2$ are integration constants which
dimensions $M^4$.
\end{enumerate}

\subsubsection{Matter solutions}

In this case, we search for $f(R)$ models which have a dust-matter
behavior, that is $a=(t/t_0)^{2/3}$,  \be \bar
H=\frac2{3\,t_0\,a^{3/2}}=\frac{H_0}{a^{3/2}}\, ,
\qquad\textrm{and}\qquad \bar
R=-\frac{2(2/t_0^2+9\,k\,a)}{3\,a^3}\, , \ee where $H_0\equiv
2/(3\,t_0)$. For the case $k=0$, we  find the explicit  analytic
solution \bea
R&=&-\frac4{3\,t_0^2\,a^3}\, ,\\
f(a)&=&a^{-(7+\sqrt{73})/4} \left(d_1
   \,a^{\sqrt{73}/2}+d_2\right)+\frac{\rho_{m0}\,a-6\rho_{r0}}{2\,a^4}\, .
\eea This is a 2-parameters family of solutions, depending on the
two integration constants $d_{1,2}$ both with dimensions $M^4$.
The Einstein-Hilbert case $f(R)=R$  belongs to this family, when
$d_1$, $d_2$, and $\rho_{r0}$ all vanish.

\subsubsection{Exponential solutions}

In this case, we look for the behavior \be \bar H=H_0={\rm
constant}\, ,\qquad\textrm{which is}\qquad \bar
R=-12\,H_0^2-\frac{6k}{a^2}\, . \ee As above, we have  three cases
depending on $k$.
\begin{enumerate}
\item  $k=0$. Both $H$ and $R$
are constants, and $R=R_0\equiv-12\,H_0^2$. The Friedmann equation
is \be
 f(R_0)-\tfrac12\,f_R(R_0)\,R_0=\frac{\rho_{m0}}{a^3}+\frac{\rho_{r0}}{a^4}\, ,
\label{infla} \ee and it has  solutions only for
$\rho_{m0}=\rho_{r0}=0$ being $R_0$ a constant (see also
\cite{barrow}).
\item  $k=1$. In this case, $H$ is still a constant but $R$ is not. One finds
\bea
R&=&-12\,H_0^2-\frac{6\,\kappa}{a^2}\,\\
f&=&d_1 \cosh\! \left(\frac{\sqrt{2 \kappa}}{H_0\,a}\right)+d_2 \sinh\!
   \left(\frac{\sqrt{2 \kappa}}{H_0\,a}\right)\notag\\
&&{}
+\frac{6 \rho_{r0}\,H_0^4}{\kappa^2}+\frac{3 \rho_{m0} \,H_0^2}{a
   \,\kappa}+\frac{6 \rho_{r0} \,H_0^2}{a^2
   \,\kappa}+\frac{\rho_{r0}}{a^4}+\frac{\rho_{m0}}{a^3}\, .
\eea
\item  $k=-1$. The solution is
\bea
R&=&-12\,H_0^2-\frac{6\,\kappa}{a^2}\, ,\\
f&=&d_1 \cos\! \left(\frac{\sqrt{-2 \kappa}}{H_0\,a}\right)+d_2 \sin\!
   \left(\frac{\sqrt{-2 \kappa}}{H_0\,a}\right)\notag\\
&&{}+\frac{6\rho_{r0}\,H_0^4}{\kappa^2}
+\frac{3\rho_{m0}\,H_0^2}{a
   \,\kappa}
+\frac{6 \rho_{r0}\,H_0^2}{a^2
   \,\kappa}
+\frac{\rho_{r0}}{a^4}+\frac{\rho_{m0}}{a^3}\, .
\eea
\end{enumerate}

\subsubsection{$\Lambda$CDM solutions}
Let us now look for  $f(R)$ models which are compatible with the
$\Lambda$CDM being solutions of Friedmann equations. This analysis
could be extremely important  to compare the $f(R)$ approach with
observations (see also \cite{leandros}).  One defines \be \bar
H^2=H_0^2\left[\frac{\Om_{m0}}{a^3}+\frac{\Om_{r0}}{a^4}+1-\Om_{m0}-\Om_{r0}\right]
.\ \ee
The differential equation to solve is
therefore the following \bea &&f''+\left[\frac{6 \Om_{m0} H_0^2}{3
\Om_{m0} H_0^2+4 a k}-\frac{4
   (\Om_{m0}+\Om_{r0}-1) a^4-7 \Om_{m0} a-8 \Om_{r0}}{-(\Om_{m0}+\Om_{r0}-1) a^4+\Om_{m0} a+\Om_{r0}}\right]\,\frac{f'}{2\,a}\notag\\
&&\qquad{}-\frac{3 \Om_{m0} H_0^2+4 a k}{2 a \left[-(\Om_{m0}+\Om_{r0}-1) a^4+\Om_{m0}
   a+\Om_{r0}\right] H_0^2}\,f\notag\\
&&\qquad{}=
-\frac{\left(3 \Om_{m0} \,H_0^2+4 a k\right) (\rho_{r0}+a \rho_{m0})}{2 a^5
   \left[-(\Om_{m0}+\Om_{r0}-1) a^4+\Om_{m0} a+\Om_{r0}\right] H_0^2}\, ,
\eea The general integral can be   numerically achieved by giving
suitable initial conditions for $f_0$, $f'_0$. This analysis will
be pursued in a forthcoming paper.

\section{Discussion and Conclusions}
\label{sec6}

In this paper, we have discussed a general method to find out
exact/analytical cosmological solutions in $f(R)$ gravity. The approach is based on
the search for Noether symmetries  which allow to reduce the
dynamics and, in principle, to solve more easily the equations of
motion. Besides, due to the fact that such symmetries are always
related to conserved quantities, such a method can be seen as a
physically motivated criterion.

The main point is that the existence of the symmetry allows to fix
the form of $f(R)$ models assumed in a point-like cosmological
action where the FLRW metric is imposed. It is worth noticing that,
starting from a point-like FLRW Lagrangian, and then deriving the
Euler-Lagrange equations of motion, leads exactly to the same
equations obtained by imposing the FLRW metric in the Einstein
field equations. This circumstance allows to search ``directly'' the
Noether symmetries in the point-like Lagrangian and then to plug
the related conserved quantities into the equations of motion. As
a result $i)$ the form of the $f(R)$ is fixed directly by the
symmetry existence conditions and $ii)$ the dynamical system is
reduced since some of its variables (at least one) is cyclic.

The method is useful not only in a cosmological context but it
works, in principle, every time a canonical, point-like Lagrangian
is achieved (in \cite{stabile}, it has been used to find out
spherically symmetric solutions in $f(R)$ gravity).

In this paper, we have considered a generic $f(R)$ theory where
standard fluid matter (dust and radiation) is present. The Noether
conditions for symmetry select  forms of $f(R)$ depending on a set
of cosmological parameters such as $\{\rho_{r0},\rho_{m0},k,H_0\}$ and
the effective gravitational coupling.  Such a dependence can be
easily translated into the more suitable set of observational
parameters $\{\Omega_{r0},\Omega_{m0},\Omega_k, H_0\}$ and then
matched with data. This situation has a twofold relevance: from
one side, it could contribute to remove the well known problem of
degeneracy (several dark energy models  fit the same data and,
essentially, reproduce the $\Lambda$CDM model); from the other
side, being the search for Noether symmetries a relevant approach
to find out conserved quantities in physics, this could be an
interesting method to select models motivated at a fundamental
level. It is worth noticing that the Noether constant of motion,
which we have found, has the dimensions of a mass and is directly
related to the various sources  present into dynamics. In some
sense, the Noether constant ``determines'' the bulk of the various
sources as $\rho_{m0}$, $\rho_{r0}$ and the effective
$\rho_{\Lambda}$ and then could greatly contribute to solve the
dark energy and dark matter puzzles. In a forthcoming paper, we
will directly compare the solutions which we have presented here
with observational data.

The ``non-Noether solutions'' deserve a final remark. In this case,
we do not ask for  a Noether symmetry but, finding these
solutions, can be related to the previous general method. We have
shown that the standard cosmological behaviors of the usual
Einstein-Friedmann cosmology can be achieved also in generic
$f(R)$ models,  assuming that the cosmological quantities $H$ and
$R$ depend on the scale factor $a$. As result, we find out general
$f(R(a))$ where the standard solutions of the linear $f(R)=R$ case
are easily recovered.
%However, these further solutions can be related to the presence of a Noether symmetry.

\bigskip
\noindent{\rm Acknowledgment}. We want to thank  prof.\ Ringeval
and prof.\ Fabri for useful discussions and comments. ADF is
supported partly by STFC, UK and partly by the Belgian Federal
Office for Scientific, Technical and Cultural Affairs through the
Interuniversity Attraction Pole P6/11. We thank also the Referee
for the fruitful discussion which allowed us to improve the paper.

\appendix

\section{Solutions and link with scalar-tensor theories}

We will explicitly show, as an example, that equation
(\ref{solzH})  is indeed a Noether solution (with $k=0$, flat
space, and $\mu_0=0$, zero Noether charge). First, from $H(a)$,
given by \bea H^2&=&-\frac{4 d_1 d_2 (-c_3)^{9/2}}{a^4}+\frac{24
d_1 d_2
   (-c_3)^{7/2}}{a^2}+\frac{\rho_{0m} d_2
   (-c_3)^{5/2}}{a^4}-36 d_1 d_2 (-c_3)^{5/2}\notag\\
&&{}+\frac{2
   \sqrt{3} \rho_{r0} {\rm arctanh}\!\left(\frac{\sqrt{3}
   a}{\sqrt{-c_3}}\right) d_2
   c_3^2}{a^4}+\frac{10 \rho_{r0} d_2
   (-c_3)^{3/2}}{a^3}
+\frac{12 \sqrt{3} \rho_{r0} {\rm arctanh}\!\left(\frac{\sqrt{3} a}{\sqrt{-c_3}}\right)
   d_2 c_3}{a^2}\notag\\
&&{}-\frac{18 \rho_{r0} d_2
   \sqrt{-c_3}}{a}+18 \sqrt{3} \rho_{r0} {\rm arctanh}\!\left(\frac{\sqrt{3} a}{\sqrt{-c_3}}\right)
   d_2\, ,\label{app1}
\eea
we can calculate the expression for $R(a)$ as follows
\bea
R&=&-12\,H^2-6\,a\,H\,H'\notag\\
&=&-\frac{144 d_1 d_2 (-c_3)^{7/2}}{a^2}+432 d_1 d_2 (-c_3)^{5/2}-\frac{48 d_2 \rho_{r0} (-c_3)^{3/2}}{a^3}-\frac{72
   \sqrt{3} d_2 \rho_{r0} {\rm arctanh}\!\left(\frac{\sqrt{3} a}{\sqrt{-c_3}}\right) c_3}{a^2}\notag\\
&&{}+
\frac{216 d_2 \rho_{r0}
   \sqrt{-c_3}}{a}-216 \sqrt{3} d_2 \rho_{r0} {\rm arctanh}\!\left(\frac{\sqrt{3} a}{\sqrt{-c_3}}\right) .\label{app2}
\eea
Since we know both $H$ and $R$, now, by using Eq.\ (\ref{genFa}), we can find $f(a)$ as follows
\be
f=-\frac{8 d_1 c_3^2}{a^3}-\frac{24 d_1 c_3}{a}-\frac{3\rho_{r0}}{a^4}+\frac{4 \sqrt{3} \rho_{r0} {\rm arctanh}\!\left(\frac{\sqrt{3} a}{\sqrt{-c_3}}\right)}{a^3 \sqrt{-c_3}}-\frac{12 \rho_{r0}}{a^2\, c_3}-\frac{12 \sqrt{3} \rho_{r0}
   {\rm arctanh}\!\left(\frac{\sqrt{3} a}{\sqrt{-c_3}}\right)}{a (-c_3)^{3/2}}\, .\label{app3}
\ee These expressions for $f, R,H$ fulfill equation (\ref{FRDA}).
The system has also a constant of motion $\mu_0=0$ given by
equation (\ref{NT2}), as the Lagrangian possesses a Noether
symmetry.

We will discuss how to link this solution  (extending this
procedure to the other solutions is straightforward) to the
scalar-tensor picture, by finding the potential for the scalar
non-minimally coupled with gravity. In fact, starting from the
action \be S=\int d^4 x\sqrt{-g}\,f(R)+S_m\, , \ee one can rewrite
it (at least at the classical level) in the following form \be
S=\int d^4x\sqrt{-g}[f_\phi\,R-V(\phi)] + S_m\, , \ee where
$V=\phi\,f_\phi-f(\phi)$, and $f_\phi=\partial f/\partial\phi$.
The classical equation of motion for $\phi$ leads to $\phi=R$. One
can make a field redefinition to bring the action in the form \be
S=\int d^4x\sqrt{-g}[-\chi\,R-V(\chi)] + S_m\, , \ee where
$\chi=-f_\phi$.

In this case we can use our solutions in order to find  $V(\chi)$,
the only unknown in the theory. One can do it as follows \bea
\chi&=&-f_\phi=-f_R=-\frac{f'}{R'}\\
V&=&\phi\,f_\phi-f=R\,f_R-f=R\,\frac{f'}{R'}-R\, , \eea where
these relations are correct on shell, i.e.\  for the solutions of
the equations of motion. Using equations (\ref{app1}),
(\ref{app2}), and~(\ref{app3}), one can write down explicitly the
potential, at least for this case, as follows \bea
V(\chi)&=&3456 d_1 d_2^3 \,\chi ^3 (-c_3)^{13/2}-10368 d_2^4 \rho_{r0} \,\chi ^4 c_3^6\notag\\
&&{}-1728 \sqrt{3} d_2^3 \rho_{r0} \,\chi ^3\, {\rm arctanh}\!\left[\frac{\sqrt{3} \left(6 d_2 \,\chi  (-c_3)^{5/2}
+\sqrt{-36 d_2^2 \,\chi ^2 c_3^5-c_3}\right)}{\sqrt{-c_3}}\right]
   c_3^4\notag\\
&&{}+1728 d_2^3 \rho_{r0} \,\chi ^3
   \sqrt{-36 d_2^2 \,\chi ^2 c_3^5-c_3} (-c_3)^{7/2}
-288 d_1 d_2 \,\chi  (-c_3)^{5/2}
+432 d_2^2 \rho_{r0} \,\chi ^2\,c_3^2\notag\\
&&{}+288 \sqrt{3} d_2^2 \rho_{r0} \,\chi ^2 \sqrt{-36 d_2^2 \,\chi ^2 c_3^5-c_3}\, {\rm arctanh}\!\left[\frac{\sqrt{3} \left(6
   d_2 \,\chi  (-c_3)^{5/2}
+\sqrt{-36 d_2^2 \,\chi ^2 c_3^5-c_3}\right)}{\sqrt{-c_3}}\right] (-c_3)^{3/2}\notag\\
&&{}+144 \sqrt{3} d_2\rho_{r0} \,\chi \, {\rm arctanh}\!\left[\frac{\sqrt{3} \left(6 d_2 \,\chi  (-c_3)^{5/2}
+\sqrt{-36 d_2^2 \,\chi ^2 c_3^5-c_3}\right)}{\sqrt{-c_3}}\right]
-16 d_1 \sqrt{-36 d_2^2 \,\chi ^2 c_3^5-c_3}\notag\\
&&{}-\frac{96 d_2 \rho_{r0} \,\chi  \sqrt{-36 d_2^2 \,\chi ^2 c_3^5-c_3}}{\sqrt{-c_3}}
-\frac{9 \rho_{r0}}{c_3^2}-576 d_1 d_2^2 \,\chi ^2 \sqrt{-36 d_2^2 \,\chi ^2 c_3^5-c_3} c_3^4\notag\\
&&{}+8 \sqrt{3} \rho_{r0}(-c_3)^{-5/2}
   \sqrt{-36 d_2^2 \,\chi ^2 c_3^5-c_3}\, {\rm arctanh}\!\left[\frac{\sqrt{3}
\left(6 d_2 \,\chi  (-c_3)^{5/2} +\sqrt{-36 d_2^2 \,\chi ^2
c_3^5-c_3}\right)}{\sqrt{-c_3}}\right] . \eea In order to study
the evolution of the background,  whether or not it leads to a viable dynamics
for the universe, it is already sufficient to check if the Hubble
parameter given by (\ref{app1}) can fit the data, from Big Bang
Nucleosynthesis up to Dark Energy domination.

\end{document}